%% file: aa55643-25_astro-ph.tex
\documentclass{aa}  

\usepackage{graphicx}
\usepackage{txfonts}
\usepackage{lipsum}
\usepackage{subcaption}         
                                
\usepackage{lscape}             
                                
\usepackage{placeins}

\usepackage{sidecap}

\def\targ{G105.42$+$9.88}

\def\nh3{NH$_{3}$}
\def\kms{km~s$^{-1}$}

\def\Vlsr{$V_{\rm LSR}$}
\def\Jyb{Jy~beam$^{-1}$}

\def\G24{G24.78$+$0.08}
\def\HII{H{\sc ii}}

\newcommand{\ms}{$M_{\odot}$}
\newcommand{\ls}{$L_{\odot}$}

\newcommand{\pas}{$\rlap{.}^{\prime\prime}$}
\newcommand{\pss}{$\rlap{.}^{\rm s}$}

\newcommand{\degree}{$^{\circ}$}

\newcommand{\et}    {et al.}

\begin{document}

   \title{Protostellar Outflows at the EarliesT Stages (POETS). VIII. The jets in the intermediate-mass star-forming region \targ\ (alias LkH$\alpha$~234)}

   \titlerunning{Jets in G105.42$+$9.88}

   \author{L. Moscadelli\inst{1}
        \and F. Massi\inst{1}
        \and O. Bayandina\inst{2,1}
        }

   \institute{INAF-Osservatorio Astrofisico di Arcetri, Largo E. Fermi 5, I-50125, Firenze, Italy
             \email{luca.moscadelli@inaf.it}
             \and
            SKA Observatory, 2 Fir Street, Observatory 7925, Cape Town, South Africa 
           }

  \abstract
   {Intermediate-mass protostars can be the preferred targets to study star formation, since they allow us to inquire how the stars form in clustered environments at a relatively close distance from us.}
   {Our aim is to investigate the formation, interaction with the local environment, and propagation of the protostellar jets inside the young stellar object (YSO) cluster \targ\ (alias LkH$\alpha$~234). This is one of the least luminous targets of the Protostellar Outflows at the EarliesT Stages (POETS) survey, which has been recently carried out to study young outflow emission on scales of 10--100~au.} 
   {We studied the jet launching regions at radii of $\le$~100~au by employing three different sets of very long baseline interferometry (VLBI) observations of the 22~GHz water masers spanning a time of $\approx$~22~yr. We complemented the VLBI maser data with sensitive multi-band NSF’s Karl G. Jansky Very Large Array (JVLA) continuum observations to image the ionized component of the outflow at scales of \ 100--1000~au and sensitive Large Binocular Telescope (LBT) H$_2$ 2.12~$\mu$m observations to track the jets at scales of 0.1--1~pc.}
   {Our combined observations allow us to study the protostellar outflows from the intermediate-mass binary system VLA~3A~and~3B, separated by $\approx$~0\pas22, and from VLA~2, an intermediate-mass YSO placed $\approx$~1$^{\prime\prime}$ to northwest of VLA~3. In each of these three YSOs, the presence of a thermal jet is suggested by previous Very Large Array (VLA) and our JVLA radio continuum observations. Toward VLA~2 the 2001 and 2011 Very Long Baseline Array (VLBA) observations consistently show that the water masers are tracing a compact (size $\approx$~400~au) bipolar collimated (PA $\approx$ 70\degr) outflow, i.e., a jet. The analysis of the 3D flow velocities proves that the jet is magneto-centrifugally launched in a magnetohydrodynamic (MHD) disk wind (DW). We infer launch radii in the range 10--50~au for the streamlines traced by the water masers. The global VLBI 2023 water maser observations indicate that the jet propagation can be hindered by a very dense clump placed northeast of VLA~2 and that is consistent with the large-scale LBT H$_2$ emission, tracing only the southwest lobe of the VLA~2 jet. Instead, the parallel (PA $\approx$ 55\degr--57\degr) jets emitted by the nearby YSOs VLA~3A~and~3B can be reliably tracked with the H$_2$ emission at scales of a few 10$^{\prime\prime}$ to both the southwest and the northeast. In particular, northeast of VLA~3 the direction of these two jets crosses a linear chain of spaced H$_2$ knots, which is a clear signature of an episodic jet. In VLA~3B the spatial distribution and intensity of the water masers change significantly between the VLBA 2011 and global VLBI 2023 observations, likely reflecting a different state (active in 2011 and quiescent in 2023) of wind ejection. The variable ejection from VLA~3B could be the origin of the episodic jet observed at larger scales.}
   {This work shows that the combination of VLBI observations of the 22~GHz water masers with sensitive high-angular-resolution radio continuum and near-infrared H$_2$ 2.12~$\mu$m observations permit to investigate the launching region of protostellar outflows, to trace their interaction with the surrounding environment, and track their paths at larger scales.}


   \maketitle

\section{Introduction}

\nolinenumbers

Observations and models indicate that the formation of stars within the mass range \ 0.1--30~\ms\  proceeds in a similar manner, through a disk-outflow system in which mass accretion and ejection are intimately related. If the basic properties of the star-forming process does not depend on the stellar mass, intermediate-mass ($\approx$1--6~\ms) protostars can be the best targets to study star formation for several reasons. They are not so rare as high-mass stars such that a high number of them can be found at a shorter distance from us, $\le$~1~kpc. Their formation time \citep[0.5--5~Myr,][]{Tes99} is still long enough to allow us to track the different phases of their evolution. With respect to low-mass star-forming clouds the environment of intermediate-mass protostars is much more clustered, containing up to a few tens of members within a distance of $\sim$~0.1~pc \citep{Tes99,Mas00,Mas03,Mos21}. That permits to observe the effects of the interactions among cluster members, which can be fundamental to determine the properties of the resulting stars.

Another advantage of the intermediate-mass protostars with respect to their low-mass siblings is that they are sufficiently powerful to sustain the excitation of intense ($\ge$~10~Jy) 22~GHz water masers over long periods \citep[up to many years as in the intermediate-mass star-forming region NGC2071,][]{Lek11}. The 22~GHz water masers are observed in low-mass protostars \citep{Fur03}, too, but they are generally much weaker and can fade away over typical monitoring periods of a few months, which makes it difficult to employ them as kinematic probes.
As indicated by previous studies  \citep{Mos07,San10b,Mos11a,God11a}, very long baseline interferometry (VLBI) observations of the 22~GHz water masers, by achieving linear resolutions of $\sim$1~au and measuring 3D velocities, can directly trace the velocity field of protostellar winds. Our recent "Protostellar Outflows at the Earliest Stages" (POETS) survey \citep{Mos16,San18} has imaged the inner portion of the wind (on scales of \ 10--100~au) in a statistically significant sample (37) of luminous young stellar objects (YSOs). We have employed multi-frequency NSF’s Karl G. Jansky Very Large Array (JVLA) observations to determine the spatial structure of the ionized emission and multi-epoch Very Long Baseline Array (VLBA) observations (from the BeSSeL\footnote{The Bar and Spiral Structure Legacy (BeSSeL) survey is a VLBA key project, whose main goal is to derive the structure and kinematics of the Milky Way by measuring accurate positions, distances (via trigonometric parallaxes), and proper motions of methanol and water masers in hundreds of high-mass star-forming regions distributed over the Galactic disk \citep{Rei14}.} survey) to derive the 3D velocity distribution of the 22~GHz water masers. During 2016--2021 about half of the POETS targets have been also observed with the Large Binocular Telescope (LBT) to image the H$_2$ 1-0 S(1) ro-vibrational line at 2.12~$\mu$m and trace the  protostellar jet emerging from the YSO at larger scales of 10$^3$--10$^4$~au (Massi \et, in preparation).

\targ, with a bolometric luminosity of $\sim$~5 $\times$ 10$^2$~\ls\ \citep{Mos16} at a distance of 0.89$\pm$0.05~kpc \citep{Xu13}, is one of the least luminous POETS targets. Near-infrared (NIR) observations by \citet{Kat11} with an angular resolution of 0\pas2  revealed a cluster of eight candidate YSOs within 10$^{\prime\prime}$ of the Herbig Be star LkH$\alpha$~234, the brightest NIR source in the region. Using the Very Large Array (VLA) \citet{Tri04} detected five radio continuum sources approximately aligned along a southeast-northwest direction within a distance of $\approx$~5$^{\prime\prime}$: LkH$\alpha$~234, VLA~1 ($\approx$~5$^{\prime\prime}$ northwest of LkH$\alpha$~234), VLA~2 ($\approx$~3$^{\prime\prime}$ northwest of LkH$\alpha$~234), and the nearby ($\approx$~0\pas22) sources VLA~3A and VLA~3B ($\approx$~2$^{\prime\prime}$ northwest of LkH$\alpha$~234). VLA~3~(A$+$B) and VLA~2 correspond to the intense NIR objects NW1 and NW2, respectively, observed by \citet{Kat11}. Another embedded source, FIRS1-MM1, was discovered in millimeter observations by \citet{Fue01} 4$^{\prime\prime}$ northwest of LkH$\alpha$~234 in between of VLA~2 and VLA~1. 22~GHz water maser emission is mainly associated with the sources VLA~2 and VLA~3B and was previously studied with both single-epoch \citep{Mar05} and multi-epoch \citep{Tor14} VLBA observations. The most interesting result from these observations was the discovery of a compact bipolar collimated outflow emerging from VLA~2 traced by the water masers \citep{Tor14}. 

Within the POETS survey, our analysis of the water masers in \targ\ was restricted to the source VLA~3 only, which is the dominant radio continuum emitter. The 22~GHz water masers toward VLA~3B exhibited a very peculiar local-standard-of-rest (LSR) velocity (\Vlsr) distribution \citep[][see their Fig.~B.7]{Mos19b}, presenting many maser clusters with a large spread in \Vlsr\ (up to $\approx$~20~\kms). In the present work, we employ the BeSSeL VLBA data to study the maser emission also toward the source VLA~2, the second most intense radio source in the region. In 2023 we have reobserved the water masers in \targ\ with sensitive global VLBI observations to detect the weakest masers and study the maser time variability.
The combination of different epochs of VLBI 22~GHz maser observations with sensitive JVLA radio and LBT H$_2$ 2.12~$\mu$m data allows us to accurately map the outflows from the sources VLA~2 and VLA~3 at length scales from 10~au to 10$^4$~au.
Sect.~\ref{Obs} describes the maser VLBI and LBT observations, and the data calibration and analysis is reported in Sect.~\ref{calana}. The combined maser VLBI and JVLA radio images at length scales of $\sim$100~au are presented in Sect.~\ref{Res}. In Sect.~\ref{Dis} we interpret the maser kinematics and discuss a coherent picture that also explains the large-scale H$_2$ emission pattern. Our main conclusions are given in Sect.~\ref{Con}.



  \begin{figure*}
    \includegraphics[width=0.48\textwidth]{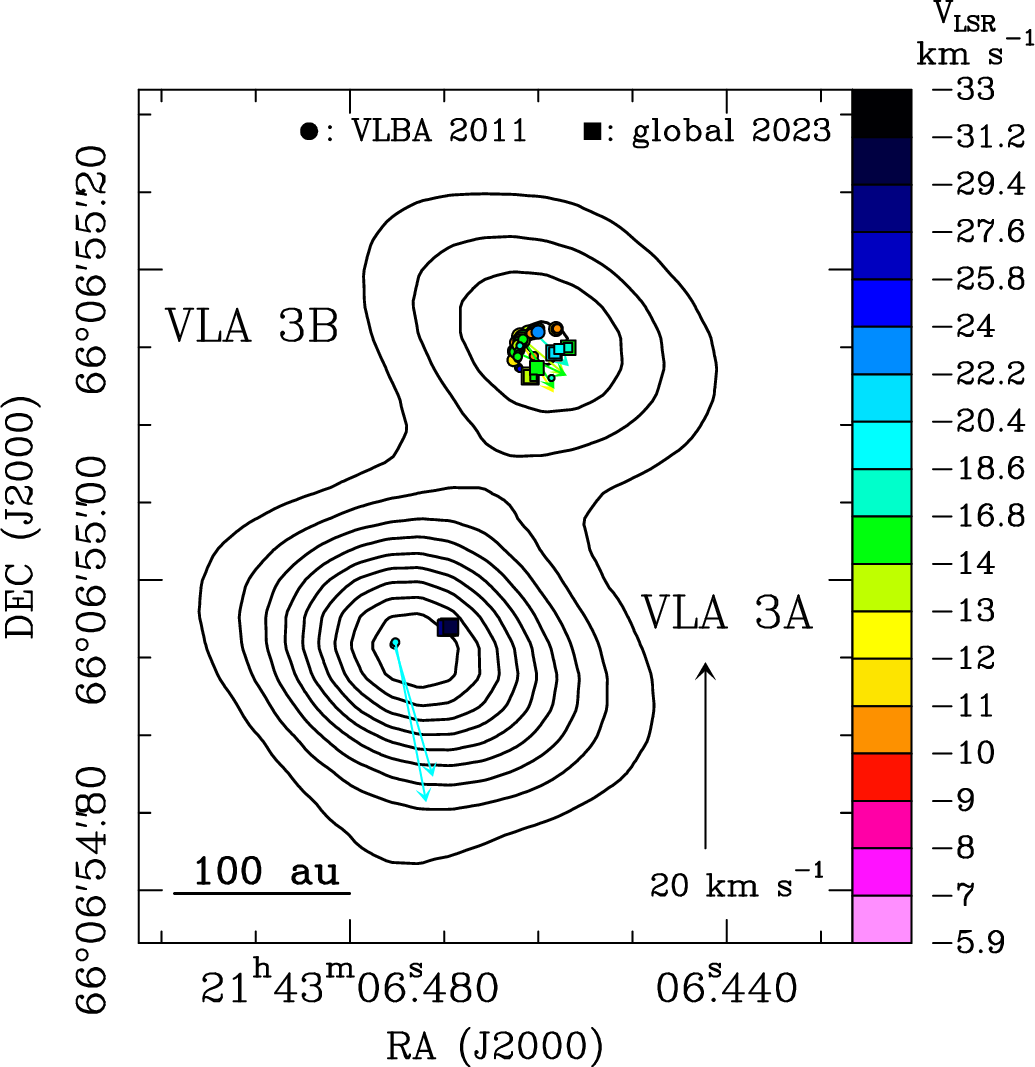} 
    \hspace*{0.3cm}\includegraphics[width=0.54\textwidth]{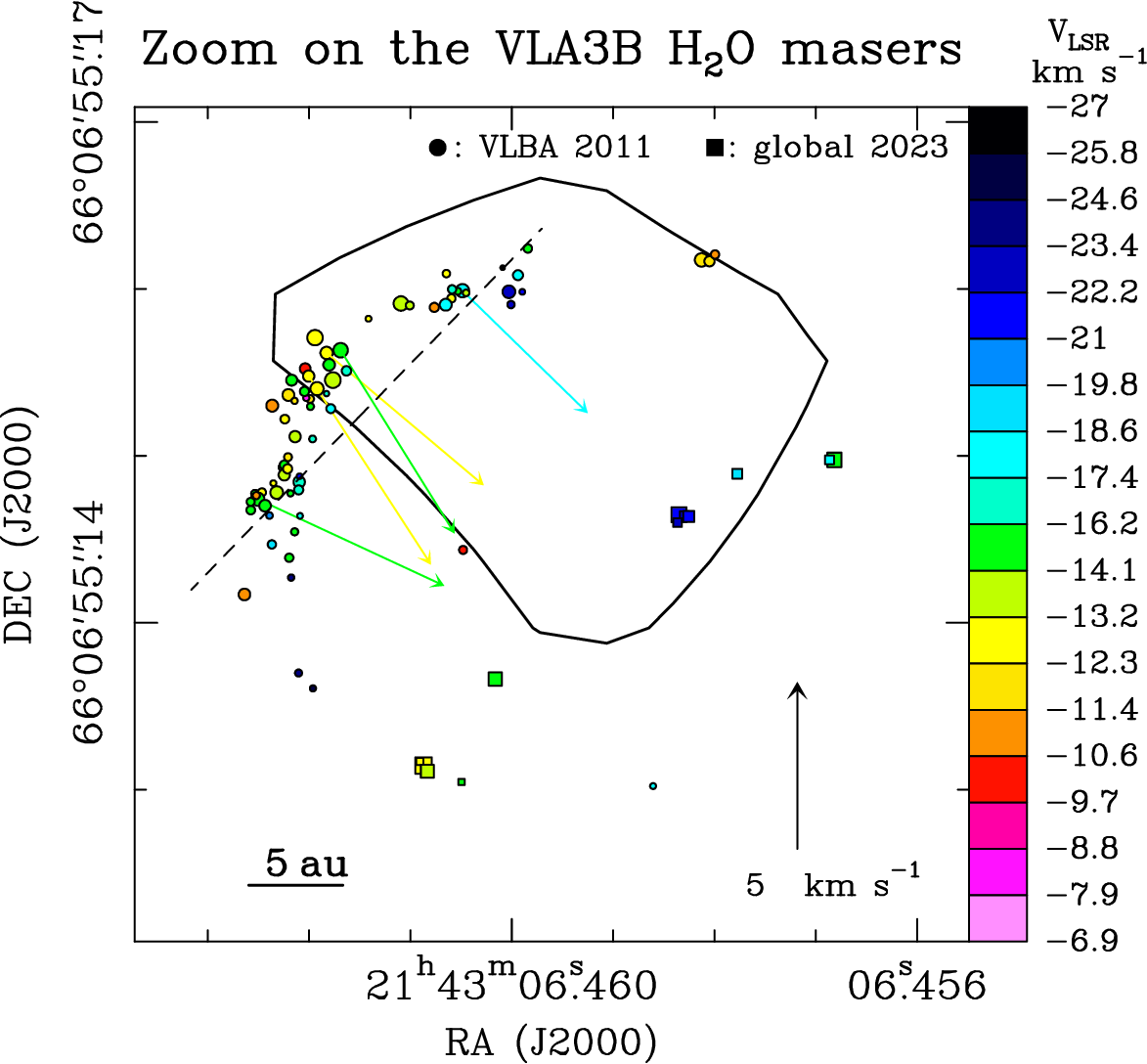}
    \caption{BeSSeL VLBA and global VLBI water masers toward VLA~3. \ (Left~panel)~Absolute positions (relative to the epoch June 6, 2023) are reported with colored dots and squares for the water masers from the BeSSeL VLBA and global VLBI  observations, respectively. Colors denote the maser \Vlsr, and masers from the same VLBI observation have symbol area proportional to the logarithm of their intensity. Colored arrows represent the absolute proper motions of the water masers measured through the six VLBA epochs, with the velocity scale shown in the bottom right of the panel. Only the proper motions with directions accurate within 45\degree\ are plotted. Black contours represent the 13~GHz JVLA A-configuration continuum observed within the POETS survey. Plotted contours are \ 10\% to 90\% (in steps of 10\%) of the map peak equal to 0.9~m\Jyb. \ (Right~panel)~Zoom on the cluster of water masers located at the peak of the VLA~3B radio continuum emission. The dashed line draws the major axis of the spatial distribution of the BeSSeL VLBA 2011 water masers.}
    \label{VLA3}
   \end{figure*}

\section{Observations}
\label{Obs}

\subsection{VLBI water maser observations} 
\label{vlbiwat}

We observed the $6_{16} - 5_{23}$ H$_2$O maser transition \citep[rest frequency 22.2350798 GHz,][]{Pick98} toward \targ\ with both the VLBA and the global VLBI array. To determine the absolute positions and velocities, we performed phase-reference observations, alternating scans on the target and the phase-reference calibrators every 22--45~s; the phase-reference calibrators were the quasars \ J2159$+$6606 (separation from the maser target of 1.65\degree, 8~GHz correlated-flux of \ $\approx$110~mJy), J2125$+$6423 (2.52\degree,  $\approx$50~mJy), and J2203$+$6750 (2.62\degree,  $\approx$120~mJy). Calibration scans of 1--10~min were observed every few hours on the fringe-finder and bandpass calibrators \ J0019$+$7327, J0102$+$5824, J0319$+$4130,  J1824$+$1044, J2202$+$4216, and J2005$+$7752,

We recorded dual circular polarization through four adjacent bandwidths of 8 and 16~MHz for the VLBA and global VLBI observations, respectively; one of the four bandwidths was centered at the target systemic \Vlsr \ of \ $-18.0$~\kms. The four bandwidths were used to increase the signal-to-noise ratio (S/N) of the weak continuum phase-reference calibrators. The VLBA  data were correlated with the VLBA-DiFX software correlator in Socorro (New Mexico, USA) and the global VLBI data with the SFXC correlator at the Joint Institute for VLBI in Europe (JIVE, in Dwingeloo, the Netherlands). For each observation we employed two correlation passes: \ 1)~256 (for VLBA) and 1024 (for global VLBI) spectral channels to correlate the maser bandwidth; and \ 2)~16 (for VLBA) and 32 (for global VLBI) spectral channels to correlate the whole set of four bandwidths. The spectral resolution attained across the maser band was 0.42~and~0.21~\kms\ for the VLBA and global VLBI observations, respectively. The correlators' averaging time was 1~s. In the following, we report the details of each of the two observations separately.

\subsubsection{BeSSeL VLBA observations}
\label{vlba}

The BeSSeL VLBA key project observed \targ\ 
(tracking center: RA(J2000) = $21^{\rm h} \, 43^{\rm m}$ 06\pss4808, Dec(J2000) = $+66$\degree\  $06^{\prime}$ 55\pas345) with the ten antennae of the  VLBA (exp. code: BR145S) of the National Radio Astronomy Observatory (NRAO\footnote{NRAO is a facility of the National Science Foundation operated under cooperative agreement by Associated Universities, Inc.}) at six epochs: May 24, August 8, October 30, November 26, 2011, and January 12, May 14, 2012.  “Geodetic” blocks \citep{Rei09} were placed before the start, in the middle, and after the end of the phase-reference observations, in order to model and remove uncompensated interferometric delays introduced by the Earth's atmosphere and improve the astrometric accuracy. Due to the different antennae available at each epoch, the full width at half maximum (FWHM) major and minor sizes of the beam, using natural weighting, vary over the observing epochs in the ranges\ 0.8--1.1~mas and 0.4--1.0~mas, respectively, and the beam position angle (PA) in the range \  [$-20$\degree, 32\degree]. In channel maps with a (relatively) weak signal, the 1$\sigma$ root mean square (rms) noise varies within \ 5--14~mJy~beam$^{-1}$.

\subsubsection{Global VLBI observations}
\label{glob}

We observed \targ\ 
(tracking center: RA(J2000) = $21^{\rm
h}  \, 43^{\rm m}$ 06\pss463,  Dec(J2000) = $+66$\degree\ $06^{\prime}$ 55\pas19) with global VLBI for 24~hr, starting on June 6, 2023, at 19:00~UT. A total of 17 antennae were involved in the observations and provided useful data: eight antennae of the European VLBI network (EVN\footnote{The European VLBI Network is a joint facility of independent European, African, Asian, and North American radio astronomy institutes.  Scientific results from data presented in this publication are derived from the following EVN project code: GM082.}),
Medicina, Jodrell\_Bank, Effelsberg, Onsala, KVN\_Ulsan, KVN\_Yonsei, Urumqi, and Tianma; plus nine antennae of the VLBA, Brewster, Fort Davis, Hancock, Kitt Peak, Los Alamos, Mauna Kea, North Liberty, Pie Town, and Saint Croix. While the EVN antennae observed the target only, the VLBA also performed phase-reference observations (over 10~hr). During the phase-reference session, the target and the fringe-finder and bandpass calibrators were always observed by the VLBA simultaneously with the EVN to ensure global VLBI baselines.  Using natural weighting, the FWHM major and minor sizes of the beam are \ 0.27~mas and 0.19~mas, respectively, and the beam PA is \  $18.0$\degree. In channel maps with a (relatively) weak signal, the 1$\sigma$ rms noise is  \ 1.3~mJy~beam$^{-1}$.

\subsection{LBT observations}
\label{obs_LBT}

We observed \targ\ with the LBT \citep{Hil06} in a program (LBT codes: LBT-2018A-C183559-15 and LBT-2018A-C183561-16, P.I.: F. Massi) aimed at imaging POETS sources
in the NIR. The observations were carried out on September 30,
2019, using the NIR imagers LUCI1 and LUCI2 \citep{Age10}
in fraternal mode.
This mode allows using both instruments simultaneously, each one matched 
to one of the
two 8.4~m mirrors of the binocular telescope and pointing toward
the same target. 
The N3.75 camera of each imager was set to imaging mode,
with a pixel scale of $\sim$~0\pas12~pix$^{-1}$ 
and a field of view of $4\arcmin \times 4 \arcmin$. The selected
filters were
the broad-band $K_{s}$ filter (centered at $2.16$~$\mu$m) and the
narrow-band H$_2$ filter (centered at the $2.12$~$\mu$m line of molecular
hydrogen). We adopted a dithering scheme whereby the 
target
was always kept inside the frame, with a total of 36
pointings in the H$_2$ band and 6 in the $K_{\rm s}$ band. The total 
on-source time, i.e., the sum of the total time spent on source 
by the two telescopes,
was $\sim$~2~hr in the H$_2$ band and $\sim$~0.2~hr in the $K_{\rm s}$ band. The photometric standard star FS143 \citep{Haw01}
was imaged through the H$_2$ band, as well,
for flux calibration. More details are given in Massi et al. 
(in preparation).

\section{Data calibration and analysis}
\label{calana}

\subsection{VLBI}
\label{VLBI}

Data were reduced with the Astronomical Image Processing System \citep[\textsc{AIPS},][]{Gre03} package following the VLBI spectral line procedures in the \textsc{AIPS} COOKBOOK\footnote{\url{http://www.aips.nrao.edu/cook.html}}. At each VLBI epoch, the emission of an intense (1--100~\Jyb) and compact maser channel was self-calibrated and the derived (amplitude and phase) corrections were applied to all maser channels before imaging. Water maser emission has been searched for over images extending  \ 2\pas0 \ in both \ RA $\cos \delta$ \ and \ DEC, and from $-$47 to 26~\kms \ in \Vlsr. We note that all the detected maser emission is found within 1\pas5 from the correlated positions, well inside the field-of-view limited by time smearing of the global VLBI and VLBA observations, which have a radius of 2\pas0 and 2\pas7, respectively. 
For a description of the criteria used to identify individual maser features, derive their parameters (position, intensity, flux, and size), and measure their absolute proper motions, we refer to \citet{Mos06}. Individual maser features are a collection of quasi-compact spots observed on contiguous channel maps and spatially overlapping (within their FWHM size). The positions of different spots are determined by fitting a two-dimensional elliptical Gaussian to their brightness distribution. For a detailed discussion of the method of calculating the error in the relative positions of spots and features, we refer to \citet{Mos24}.

Tables~\ref{wat1}~and~\ref{wat2} report the parameters (intensity, \Vlsr, position, and absolute proper motion) of the 22~GHz water masers for the source VLA~2 from the BeSSeL VLBA observations and for both VLA~2 and VLA~3 from the global VLBI observations, respectively. The 22~GHz maser parameters for the sources VLA~3A and VLA~3B from the BeSSeL VLBA observations have been already reported by \citet[][see their Tables~B.13~and~B.14]{Mos19b}.
For both the VLBA and global VLBI observations and at each epoch, inverse phase referencing \citep{Rei09} produced good (S/N $\ge$~10) images of the phase-reference calibrators. Taking into account that the calibrators are relatively compact, with sizes of $\lesssim$~1~mas, and that the absolute position of the calibrators is known within a few 0.1~mas, we estimate that the error on the absolute position of the masers is \ $\lesssim$~0.5~mas. 
Correcting for the uncompensated delay introduced by the Earth's atmosphere \citep{Rei09}, our VLBA observations reach an astrometric accuracy in the maser-referenced calibrator images of \ $\lesssim$~0.05~mas, and the absolute proper motions have an average error of \ 1.2~\kms.  
The derived absolute proper motions were corrected for the apparent motion due to the combination of the Earth's orbit around the Sun (parallax), the solar motion, and the differential Galactic rotation between our LSR and that of the maser source. We adopted a flat Galaxy rotation curve ($R_0 = 8.33\pm0.16$~kpc, $\Theta_0 = 243\pm6$~\kms\ \citep{Rei14}) and the solar motion $U = 11.1^{+0.69}_{-0.75}$, $V = 12.24 ^{+0.47}_{-0.47}$, and $W = 7.25^{+0.37}_{-0.36}$~\kms\ by \citet{Sch10}, who revised the Hipparcos satellite results.

\subsection{LBT}
\label{cal_LBT}

Data reduction was carried out using standard IRAF routines and includes
flat-fielding, bad-pixel removal, and sky subtraction. For each frame,
sky images were obtained by median-averaging 4 nearby frames. 
Finally, we summed up all the sky-subtracted frames taken through
the same filter, after
registering them, regardless of the employed telescope mirror. As LUCI1 and LUCI2
have slightly different pixel scale, we regridded the LUCI2
frames to match the LUCI1 ones using the IRAF routines \textsc{geomap} and \textsc{geotran}. The seeing-limited angular resolution in our final combined image is $\sim$~0\pas8.

In order to obtain an image of pure H$_2$ $2.12$~$\mu$m line emission,
we used the broad-band $K_{\rm s}$ image to estimate the continuum
contribution in the H$_2$ filter. We performed aperture photometry of
a large number of common stars in the $K_{\rm s}$-H$_2$ pair of combined
images using the package \textsc{daophot} in IRAF. From the
average flux ratio of H$_2$ to $K_{\rm s}$, we estimated a scaling factor for the
H$_2$ and Ks images (in the 0-order approximation that the continuum
stellar fluxes are constant in the two filters) and subtracted the scaled
images from each other, after registering them. The final subtracted
image was flux-calibrated using photometry of the standard star FS143.

To compare the location of the H$_2$ emission knots and
that of the compact radio sources VLA~3~(A$+$B) and VLA~2, 
the pure H$_2$ line emission image was astrometrically
calibrated by matching the stars in the original H$_2$ image and the GAIA DR3
catalog \citep{Gai23} using the task \textsc{ccxymatch}
in IRAF. A plate solution was obtained with the IRAF task \textsc{ccmap},
and the image header was updated accordingly. 
The absolute coordinates are in the
ICRS reference frame and should be accurate within $\sim$~0\pas1. 
As the astrometric accuracy of our radio data is much
better, $\sim$~0\pas1 is effectively the error on the relative positions of NIR versus radio sources.



  \begin{figure*}
   \sidecaption
    \includegraphics[width=12cm]{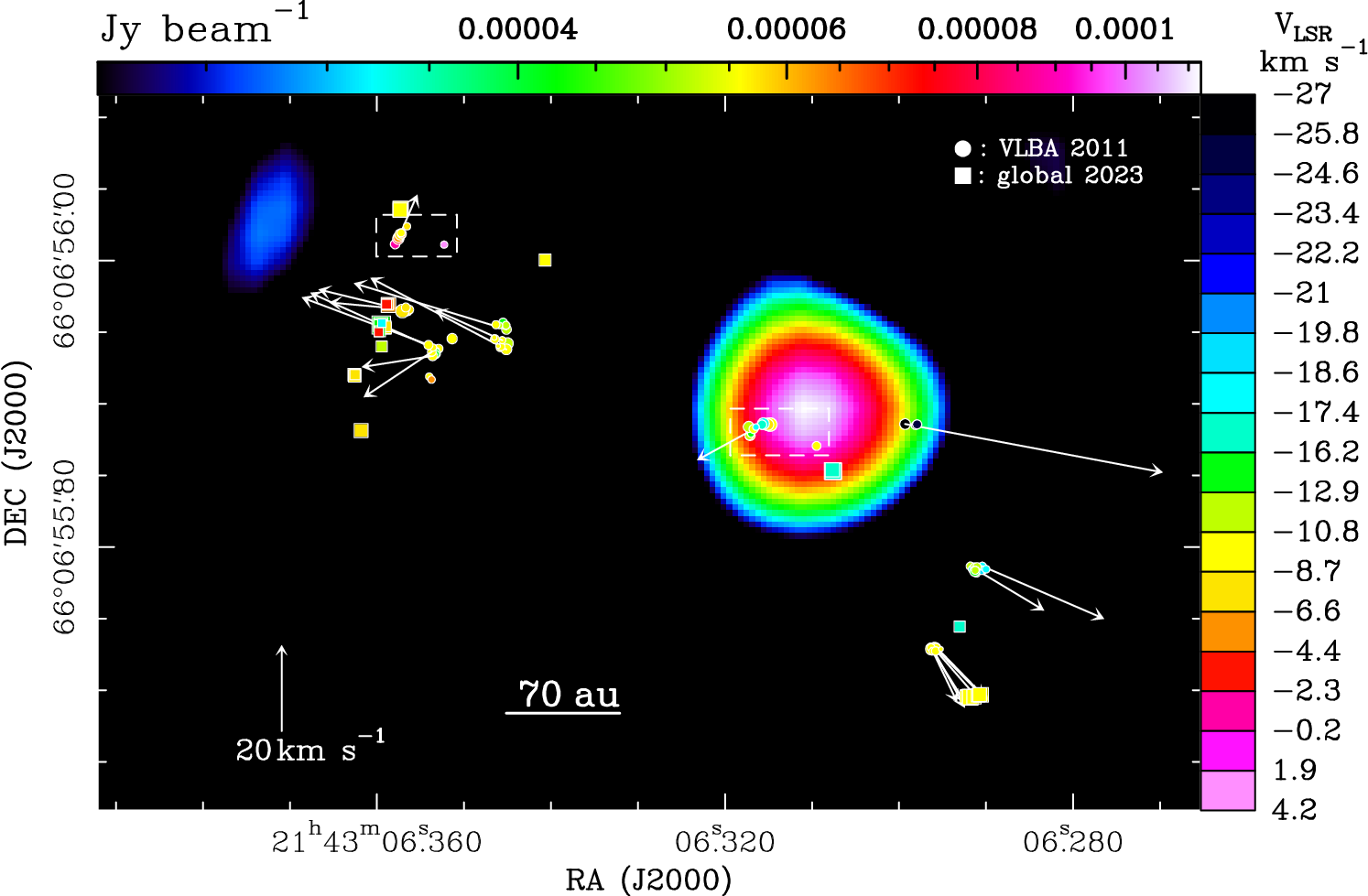} 
    \caption{BeSSeL VLBA 2011 and global VLBI 2023 observations of the water masers toward VLA~2. Colored dots and squares have the same meaning as in Fig.~\ref{VLA3}. White arrows represent the absolute proper motions of the water masers measured through the six VLBA epochs, with the velocity scale shown in the bottom left of the panel. Only the proper motions with directions accurate within 45\degree\ are plotted. The two dashed white rectangles encompass the masers that are excluded from the analysis in Sect.~\ref{VLA2-DW} (see text). The color map is the 13~GHz JVLA A-configuration continuum observed within the POETS survey, whose intensity scale is reported in the wedge on top of the panel.}
    \label{VLA2}
   \end{figure*}


  \begin{figure*}
  \sidecaption
    \includegraphics[width=12cm]{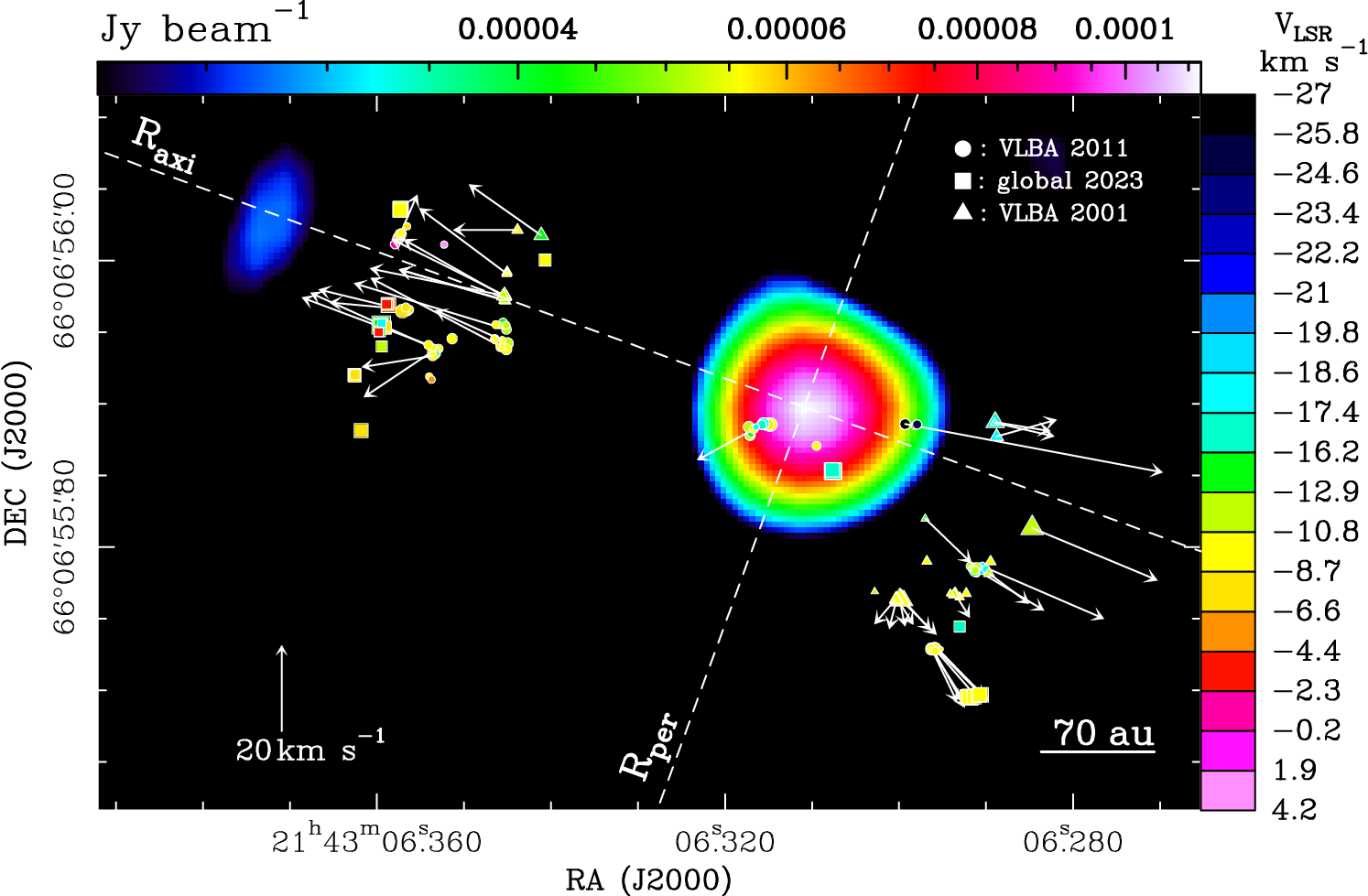} 
    \caption{VLBA 2001 \citep{Tor14}, BeSSeL VLBA 2011, and global VLBI 2023 observations of the water masers toward VLA~2. Colored dots and squares have the same meaning as in Fig.~\ref{VLA3}. Colored triangles report the absolute positions of the water masers observed with three epochs of VLBA observations by \citet{Tor14}. White arrows represent the absolute proper motions of the water masers, with the velocity scale shown in the bottom left of the panel. Only the proper motions with directions accurate within 45\degree\ are plotted. The color map is the same as in Fig.~\ref{VLA2}. The two dashed white lines indicate the jet direction, labeled $R_{\rm axi}$, and the perpendicular to the jet, $R_{\rm per}$.}
    \label{VLA2_Tor}
   \end{figure*}

\section{Results}
\label{Res}

\subsection{\targ\ VLA3}
\label{Res1}

Figure~\ref{VLA3} compares the distributions of the 22~GHz water masers toward VLA~3 from the BeSSeL VLBA 2011 and global VLBI 2023 observations. 
The results of the BeSSeL VLBA observations toward VLA~3 have been already reported by \citet[][see their Fig.~B.7]{Mos19b}.
We have corrected maser and radio continuum positions for the apparent motion between the different observing epochs. This correction ensures that the relative positions of different maser observations are accurate within 5~mas. Toward VLA~3A the VLBA and global VLBI observations are consistent, both reporting a few blueshifted masers close to the radio continuum peak. Instead, toward VLA~3B the two VLBI observations evidence a clear change in the maser properties across a time interval of $\approx$~12~yr. The elongated pattern drawn by the 2023 masers is about parallel to that of the 2011 masers but it is shifted $\approx$~20~au to the southwest, along a direction similar to that of the proper motions of the 2011 masers. Besides, the intensity of the 2023 masers, in the range 0.01--0.33~\Jyb\ (see Table~\ref{wat2}), is much smaller than that of the 2011 masers, in the range 0.1--100~\Jyb\ \citep[][see their Table~B.14]{Mos19b}. That explains the much smaller number of detection of the global VLBI observations despite the ten times higher sensitivity.

\subsection{\targ\ VLA~2}
\label{Res2}

Figure~\ref{VLA2} reports the absolute positions and proper motions of the 22~GHz water masers toward VLA~2 from the BeSSeL VLBA observations overlaid on the 13~GHz JVLA A-configuration continuum from POETS. The water masers present a bipolar spatial and velocity distribution: the maser positions are stretched along a northeast-southwest direction crossing the continuum peak and the maser proper motions are about parallel to this direction and recede from the continuum peak. Figure~\ref{VLA2_Tor} shows that the water maser velocity pattern of the BeSSeL VLBA 2011 observations is well consistent with that of the VLBA 2001 observations by \citet{Tor14}, who have interpreted the maser velocity distribution in terms of a compact bipolar collimated outflow emitted by the YSO VLA~2. In agreement with the POETS results \citep{Mos19b}, the JVLA A-configuration 13~GHz continuum peak is a good proxy for the YSO position. Very weak continuum emission\footnote{This weak source is not detected at 6~and~22.2~GHz in the POETS JVLA observations (see Sect.~\ref{VLA2-DW}), which prevents us from determining its spectral index.} is also detected along the outflow direction beyond the cluster of water masers that trace the northeast lobe of the outflow (see Fig.~\ref{VLA2_Tor}). Figures~\ref{VLA2}~and~\ref{VLA2_Tor} also report the position of the water masers from the global VLBI 2023 observations. Most of the 2023 masers are placed to the northeast of the YSO and concentrate in a linear structure at intermediate position between the 2001 and 2011 masers and the very weak continuum emission.


  \begin{figure*}
  \sidecaption
    \includegraphics[width=0.35\textwidth]{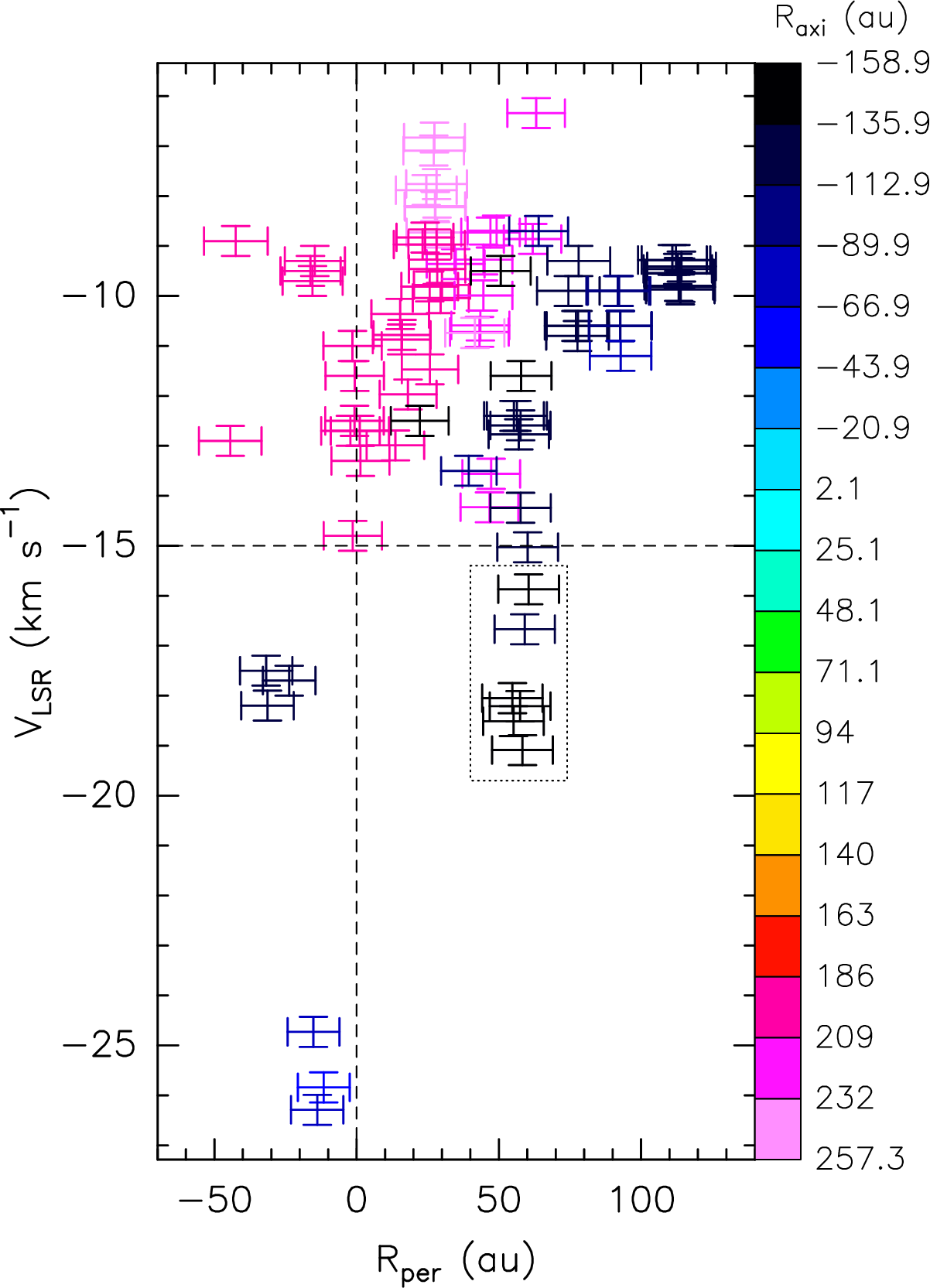} 
    \hspace*{0.3cm}\includegraphics[width=0.334\textwidth]{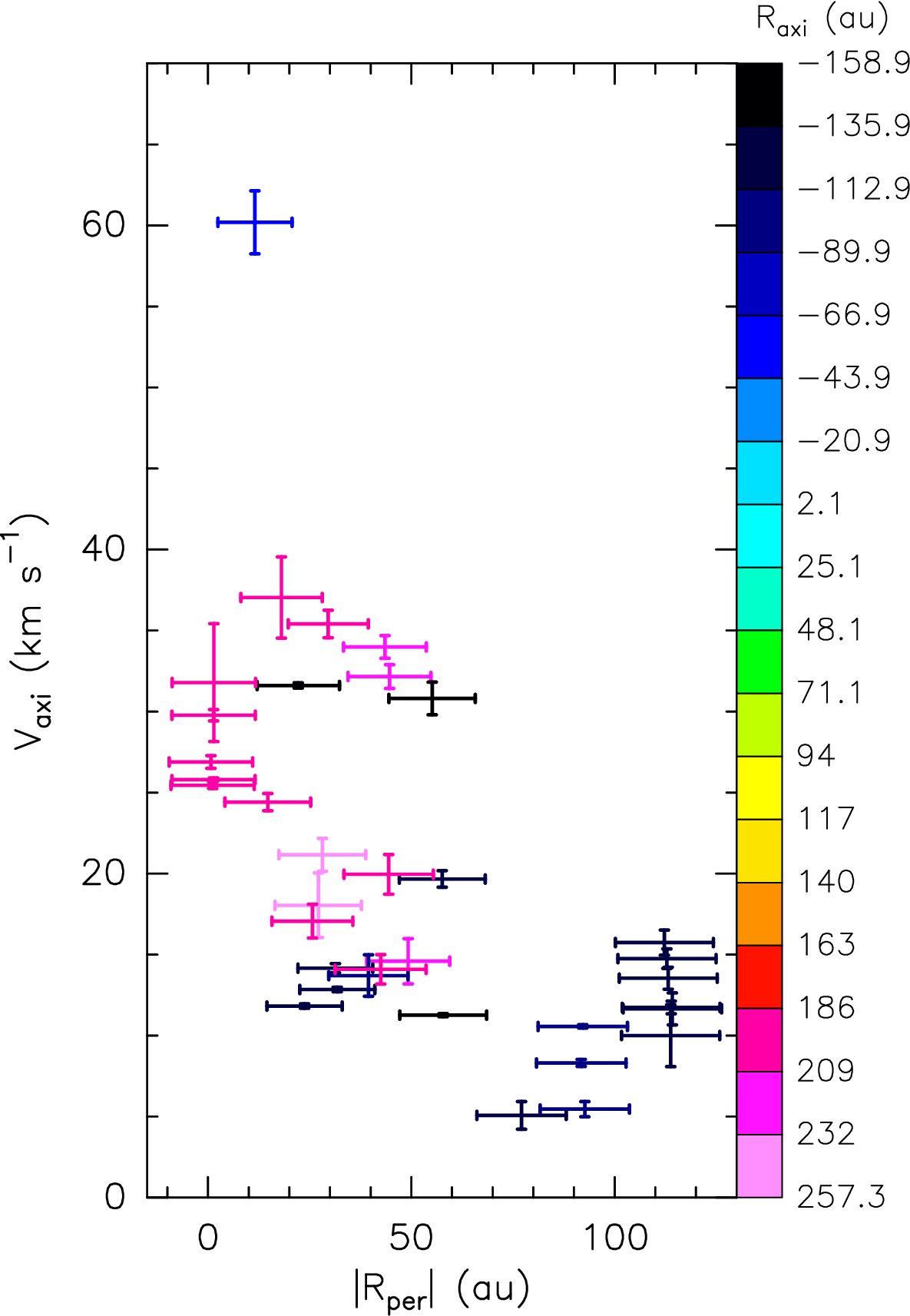}
    \caption{Spatial distribution of the wind velocities. \ (Left~panel)~Plot of the maser \Vlsr\ versus positions projected perpendicular to the jet axis, $R_{\rm per}$ (taken positive toward PA = 160\degr). The colors of the error bars indicate the position projected along the jet axis, $R_{\rm axi}$ (taken positive toward PA = 70\degr), as denoted by the wedge to the right of the panel. The two dashed vertical and horizontal lines mark the position and \Vlsr\ of the YSO, respectively.  The dotted rectangle delimits the water masers with blueshifted \Vlsr\ at positive $R_{\rm per}$. \ (Right~panel)~Plot of the maser velocity component parallel to the jet, $V_{\rm axi}$, versus the absolute value of  $R_{\rm per}$. Colors have the same meaning as in the left panel.}
    \label{V-r}
   \end{figure*}

\section{Discussion}
\label{Dis}

\subsection{The VLA~2 jet}
\label{VLA2-J}

\subsubsection{The MHD DW powering the VLA~2 jet}
\label{VLA2-DW}

The continuum source \targ\ VLA~2 has been analyzed by \citet[][see their Table~A.1 and Fig.~A.2]{San18} within the POETS survey, which has observed with the JVLA A-configuration at C (6~GHz) and Ku (15~GHz) bands and with the B-configuration at K (22.2~GHz) band. VLA~2 is not resolved at C-~and~K-bands with a beam of 0\pas3 and only slightly resolved at Ku-band with a beam of 0\pas1. Its flux increases from C-band, 53~$\mu$Jy, to K-band, 288~$\mu$Jy, with a positive spectral index equal to \ 1.3$\pm$0.3 consistent with emission from a thermal wind or jet. The extrapolated flux at 3.6~cm is 80~$\mu$Jy, slightly lower than the 4$\sigma$ estimate of 0.1~mJy by \citet{Tri04}. From the correlation between the radio continuum and bolometric luminosity for YSOs emitting a thermal jet \citep[][see their Fig.~8]{Ang18}, we infer a bolometric luminosity of a few 10~\ls\ for VLA 2, which can be provided by an embedded protostar of a few solar masses \citep[see, for instance,][]{Bre12}. In the following, we adopt a mass of \ 2$\pm$1~\ms\ for the YSO VLA~2.

Looking at Fig.~\ref{VLA2_Tor}, the collimated direction of most of the water maser proper motions and their high amplitudes, up to 40--60~\kms, hint at a magneto-centrifugal launching mechanism of the protostellar outflow. In fact, acceleration due to (only) magnetic or thermal pressure produces typical terminal flow speeds  $< 20$~\kms\ \citep{OK23b}. We consider the case that the water masers are tracing the collimated portion of a magnetohydrodynamic (MHD) disk wind (DW), i.e., the jet. The collimated maser velocities indicate that the jet axis has a PA of $\approx$~70\degr\ (denoted by the dashed white line labeled $R_{\rm axi}$ in Fig.~\ref{VLA2_Tor}). Besides, the bipolar spatial and velocity distributions of the water masers suggest that the jet axis is close to the plane of the sky. We can assume that the velocity component of the wind parallel to the jet axis,  $V_{\rm axi}$, is the dominant poloidal component and that the maser \Vlsr\ mainly reflects the toroidal component of the wind. Figure~\ref{V-r} shows the spatial distributions of the toroidal and poloidal velocity components along and across the jet axis. In producing these plots, we have employed only the VLBA 2001 and 2011 observations providing the 3D maser velocities. We have excluded the two groups of VLBA 2011 masers delimited by dashed rectangles in Fig.~\ref{VLA2}, because one group is found too close to the YSO (assumed to be placed at the peak of the radio continuum) to trace the jet and the other group presents kinematic properties more similar to those of the global VLBI 2023 masers as discussed in Sect.~\ref{Jsho}. The YSO \Vlsr\ is taken to be \ $\approx$~$-15$~\kms, which corresponds to the center of the maser velocity interval. 

A distinctive feature of a MHD~DW is that the gas rotates around the jet axis and one would expect redshifted and blueshifted gas on either sides of the jet axis. Figure~\ref{V-r} (left~panel) shows that this condition is verified for the vast majority of the water masers, which have redshifted and blueshifted velocities at positive and negative values of $R_{\rm per}$, respectively. The few exceptions can be explained by considering that the water masers trace shocks along the outflow and sometimes their motion does not reliably trace that of the flow. As an example, all the blueshifted masers at positive $R_{\rm per}$ (delimited by a dotted rectangle in Fig.~\ref{V-r}, left~panel) belong to the same maser cluster (size <~10~au), which has a large spread in \Vlsr\ ($\approx$~8~\kms) that is very likely due to local effects. 
If one considers separately the masers in the two lobes of the outflow (i.e., the blue-black and red-pink error bars in Fig.~\ref{V-r}, left~panel), there is also a clear trend for an increase in \Vlsr\ with  $R_{\rm per}$. Since a linear change in \Vlsr\ with $R_{\rm per}$ \ is expected if the masers trace an edge-on rotating ring, the significant spread present in the distribution of \ \Vlsr\ versus \ $R_{\rm per}$ \ suggests that the masers rotate in several rings. Given that most of the water masers concentrate within two small ranges of \ $R_{\rm axi}$, [$-$90, $-$130]~au \ and \ [185, 230]~au for the southwest and northeast masers, respectively, in each lobe of the outflow the rings traced by the masers should be found close in space, probably at the position where the gas physical conditions are more favorable for exciting the water masers.

Another characteristic kinematic feature of a MHD DW is that the gas closer to the jet axis moves faster. In fact, this gas is magneto-centrifugally launched at smaller disk radii and higher rotational velocities, and it also collimates at shorter distance from the disk. Fig.~\ref{V-r} (right~panel) shows that the maser flow velocities, $V_{\rm axi}$, effectively decrease with the rotation radius, $R_{\rm per}$, from 60~\kms\ to 10~\kms\ going from \ $\approx$~10~au to $\approx$~100~au. In producing this plot we have excluded a few (4 out of 39) non-collimated maser velocities, forming an angle $\ge$~45\degr\ with the jet axis. If, following our previous considerations, the masers trace rotating rings, since the velocity field of a MHD DW is axisymmetric, we should observe similar axial velocities at different sky-projected rotation radii, which can explain the observed spread in \ $R_{\rm per}$. 

The terminal flow velocity of a MHD DW can be approximated as \citep[][see Eq.~12]{Pud07}:
\begin{equation}
\label{eq_vk}
 V_{\rm axi} \approx \sqrt{2} \; \; \frac{r_{\rm A}}{r_0} \; \; \varv_{\rm K,0}
,\end{equation}
 
\noindent where \ $r_{\rm A}$ \ is the distance from the rotation axis at which the wind velocity equals the  Alfv\'{e}n speed, and $\varv_{\rm K,0}$ \ is the disk Keplerian velocity at the radius \ $r_0$. Eq.~\ref{eq_vk} holds when the flow is re-collimated along the jet axis, i.e., when \ $V_{\rm axi}$ \ is significantly greater than the toroidal velocity component. Since numerical and theoretical models \citep{Pud07,Pud19} supported by observations \citep{Bac02} constrain the ratio of \ $r_{\rm A} / r_0 \approx$~3 and the YSO mass is known, Eq.~\ref{eq_vk} allows us to estimate the launch radius \ $r_0$ \ directly from the measurement of  \ $V_{\rm axi}$. Taking the highest flow velocities \ $V_{\rm axi} \ge 20$~\kms, which are certainly magneto-centrifugally accelerated, and selecting those that are at least five times larger than the corresponding toroidal component, we infer launch radii for the maser streamlines in the range of \ 10--50~au.  We note that \citet{Mos24}, by employing water maser VLBI observations, too, have determined a similar range of launch radii for the MHD DW associated with the intermediate-mass YSO IRAS~21078$+$5211.


  \begin{figure*}
    \includegraphics[width=\textwidth]{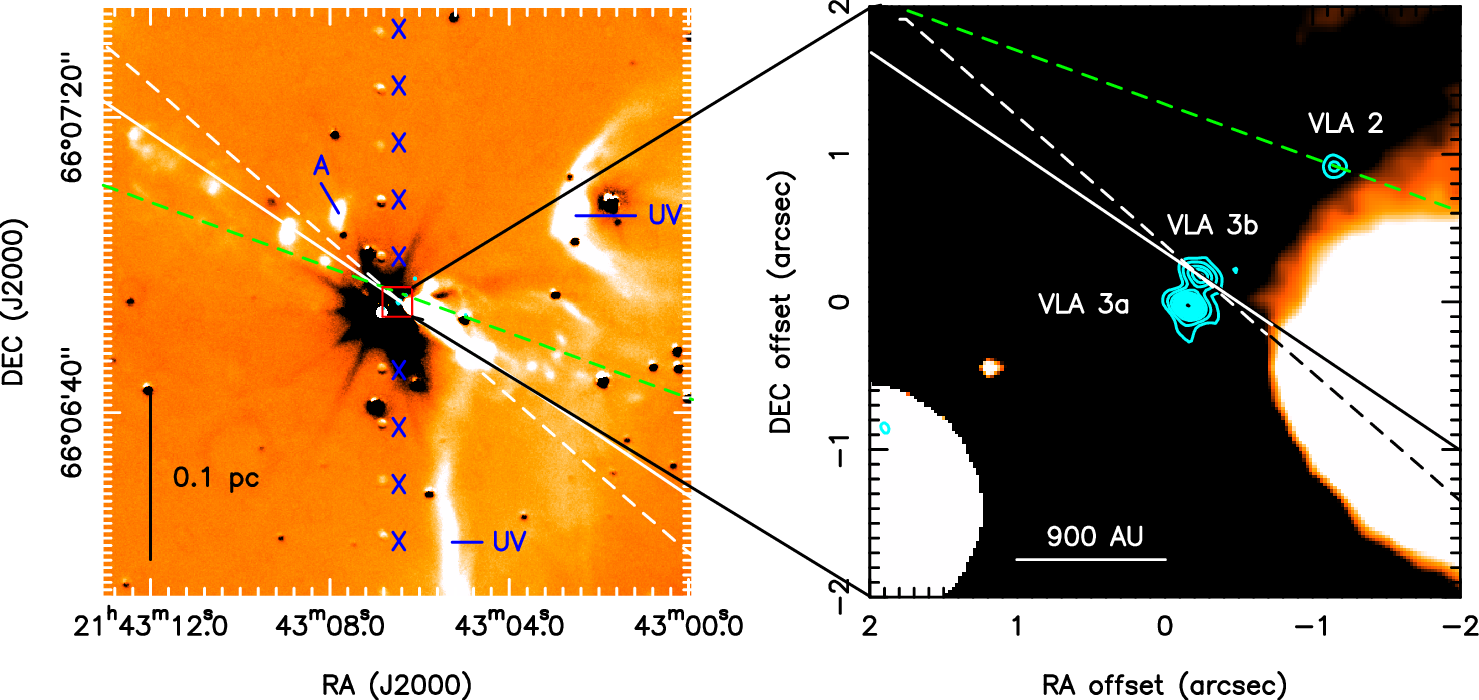} 
    \caption{H$_2$ 2.12~$\mu$m LBT observations toward the \targ\ (LkH$\alpha$~234) YSO cluster. \ (Left~panel)~Pure H$_2$ line emission (color map) overlaid with the JVLA A-configuration 13~GHz continuum (cyan contours) observed by \citet{Mos19b}. The black patches are artifact from the image subtraction due to the fact that the $K_s$-band filter used to estimate the continuum contribution of the YSO LkH$\alpha$~234 is centered at a slightly longer wavelength than the H$_2$ narrow-band filter. The emission patches south and northwest of LkH$\alpha$~234 labeled as "UV" are due to gas excited by the ultra-violet (UV) radiation of a nearby \HII\ region. The blue crosses mark artifacts due to source saturation. The knot labeled "A" corresponds in position with the H$_2$ knot "SP3" studied by \citet[][see their Fig.~10]{Oh18}. The solid and dashed white lines mark the directions of the elongated radio continuum and water maser proper motions in VLA~3B, respectively, and the dashed green line gives the direction of the water maser jet in VLA~2. \ (Right~panel)~Zoom-in on the VLA~3 and VLA~2 radio sources. Plotted contours are \ 4\%, and 10\% to 90\% (in steps of 10\%) of the map peak equal to 0.9~m\Jyb. }
    \label{H2-all}
   \end{figure*}


  \begin{figure}
    \includegraphics[width=0.5\textwidth]{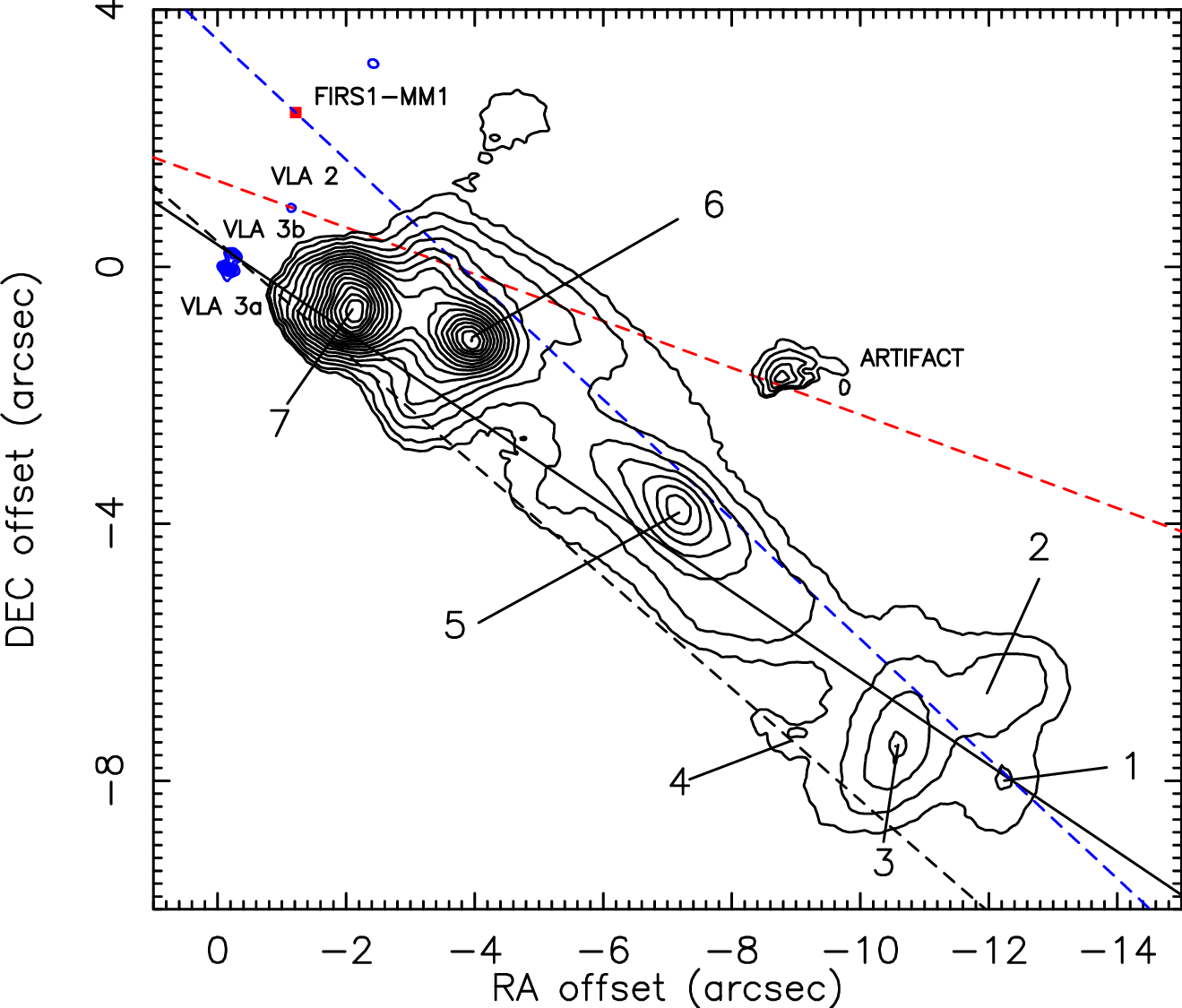}
   \caption{Contour map of pure H$_2$ 2.12~$\mu$m emission toward \targ. The black contours give the H$_2$ line emission reporting the levels at 30$\sigma$ and from 50$\sigma$ in steps of 30$\sigma$. The blue contours have the same meaning as the cyan contours in Fig.~\ref{H2-all}. The position of the millimeter source FIRS1-MM1 is indicated by a red square. The solid and dashed black lines mark the directions of the elongated radio continuum and water maser proper motions in VLA~3B, respectively, the dashed red line gives the direction of the water maser jet in VLA~2, and the dashed blue line denotes the direction of the H$_2$ jet from FIRS1-MM1 \citep{Oh18}. The main H$_2$ knots are labeled after Massi \et\ (in preparation).}
   \label{H2-SW}
   \end{figure}

\subsubsection{Local interaction of the VLA~2 jet}
\label{Jsho}

In Sect.~\ref{VLA2-DW} we have used the 2001 and 2011 VLBA maser observations together to describe the velocity field of the jet from the YSO VLA~2. This approach is justified because the two observations present a very similar velocity distribution. Now, we wish to study the time evolution of the jet by employing the three available sets of VLBI maser data (two VLBA and one global VLBI) covering a time span of $\approx$~22~yr (see Fig.~\ref{VLA2_Tor}).
We have calculated the mean value (and corresponding standard deviation) of $R_{\rm axi}$ separately for the masers in the northeast and southwest lobes of the jet. For the southwest and northeast lobe we find \ $-$115$\pm$20 and 199$\pm$4~au, $-$124$\pm$22 and 218$\pm$24~au, and  \ $-$158$\pm$7 and 262$\pm$17~au, for the 2001, 2011, and 2023 observations, respectively. In doing this calculations, we have excluded a few masers that very likely do not trace the jet, the ones from the VLBA 2011 observations (within the two dashed rectangles in Fig.~\ref{VLA2}) already excluded from the analysis of Sect.~\ref{VLA2-DW}, and the few global VLBI 2023 masers observed close ($R_{\rm axi} \approx$~$-$30~au) to the YSO (see Fig.~\ref{VLA2}).  The derived values of $R_{\rm axi}$ indicate that, in both jet lobes, the average maser position gets farther from the YSO with time, only by a small amount of 10--20~au between 2001 and 2011, by a larger separation of \ 30--40~au between 2011 and 2023.
Water masers are excited in J-~or~C-shocks at the positions where the fast gas of the jet hits local condensations of sufficiently high     
number density in the range of \ 10$^6$--10$^8$~cm$^{-3}$ \citep{Hol13}. It is plausible that the surrounding dense gas is blown by the jet to larger distances from the YSO in the course of time, which can explain the observed trend in maser position. Besides, the observed smaller displacement at earlier times agrees with the expectation to find denser gas (more efficient in braking the expansion of the jet) closer to the YSO. 

As noted in Sect.~\ref{Res2}, most of the 2023 masers are found in a linear structure facing the very weak continuum emission placed along the jet axis (see Fig.~\ref{VLA2_Tor}). This linear distribution is peculiar with respect to the less ordered and more clustered distributions of the water masers in the previous two epochs. We interpret the very weak continuum as emission from the shock-ionized portion of a very dense clump hit by the fastest gas of the jet core, which has been able to penetrate deeper into the clump. Then the linear distribution drawn by the 2023 masers could mark the shock front (seen edge-on) where the slower and more external gas of the jet hits the surface of the clump. In line with this interpretation, the maser \Vlsr\  in the 2023 linear structure is scattered and strongly deviates from that expected for jet rotation (see Fig.~\ref{VLA2_Tor}), suggesting a strong interaction of the jet with the local gas. The group of 2011 masers aligned with the 2023 masers (inside the dashed rectangle to the northeast of the YSO in Fig.~\ref{VLA2}) also presents anomalous very redshifted \Vlsr, and the measured proper motion for one maser is directed about perpendicular to the expected jet motion, hinting again at strong kinematic perturbations. In Sect.~\ref{H2} we investigate further the propagation of the VLA~2 jet using the LBT H$_2$ 2.12~$\mu$m observations.

\subsection{Large-scale view of the jets from LBT H$_2$ observations}
\label{H2}

The VLA A-configuration 8~and~22.2~GHz observations of \citet{Tri04} have resolved the nearby sources VLA~3A~and~3B allowing them to determine the structure and spectral index of each of the two radio sources. The highly elongated structures, at similar PA: 55\degr\ and 57\degr\ for VLA~3A~and~3B, respectively,  and the derived spectral indices,  1.11$\pm$0.16 and 0.69$\pm$0.29 for VLA~3A~and~3B, suggest that each of the two continuum emissions traces a thermal jet. Figure~\ref{H2-all} presents the H$_2$ 2.12~$\mu$m emission toward the LkH$\alpha$~234 YSO cluster observed with LBT at an angular resolution of 0\pas8. Two H$_2$ emission patterns, sharing a similar southwest-northeast orientation, are recognizable close to LkH$\alpha$~234, the black patch at the center of the field-of-view. To the northeast of LkH$\alpha$~234 one observes a linear chain of well-separated H$_2$ emission knots. A continuous pattern of H$_2$ emission of approximately rectangular shape is found southwest of the YSOs VLA~3~(A~and~B)~and~VLA~2. It has a length of $\approx$15$^{\prime\prime}$ and a transversal width of $\approx$3$^{\prime\prime}$. The position and well-defined orientation of the H$_2$ emission clearly indicate that it is excited by the jets emerging from the YSOs VLA~3A~and~3B, and VLA~2. In the following, we discuss in more detail the correspondence between the patterns of H$_2$ emission and each of the three jets.

A linear chain of spaced H$_2$ knots is a clear signature of a jet characterized by episodic emission. The strongest knots more detached from LkH$\alpha$~234 are equally spaced by $\approx$12$^{\prime\prime}$. Taking a jet speed of $\sim$~500~\kms\ \citep[adequate for intermediate-mass YSOs,][]{Tor14} the observed knot spacing corresponds to a travel time of  $\sim$170~yr (at the source distance of 0.89~kpc), which can be an estimate for the time separation between consecutive episodes of emission. The extent of each knot along the jet direction is only a few arcsec, which suggests that each episode of emission lasts $\sim$30--50~yr.
As shown in Fig.~\ref{H2-all}, the direction of the parallel jets from the YSOs VLA~3A~and~3B crosses the chain of H$_2$ knots, which indicates that this pattern of H$_2$ emission traces the northeast lobe of these jets at larger scales. Toward VLA~3A the VLBA 2011 and global VLBI 2023 observations are consistent, both recovering only a few weak water masers. Instead, as commented in Sect.~\ref{Res1}, the water masers associated with VLA~3B are highly variable on short timescales of $\approx$12~yr.
In 2011 the intense water masers are distributed along a direction that is about perpendicular to that of the jet and could be tracing a series of shocks all over the disk surface from where the MHD DW powering the jet is being launched. In 2023, on the contrary, we observe only a few weak water masers detached from the (putative) disk, suggesting that the episode of wind ejection is terminated. The highly variable ejection from VLA~3B revealed by the water masers at scales of $\sim$~10~au could be the origin of the episodic jet traced by the northeast chain of H$_2$ knots at scales of  $\sim$~0.1~pc.

Figure~\ref{H2-SW} zooms on the structure of the H$_2$ emission pattern extending to southwest of VLA~3, VLA~2, and the millimeter source FIRS1-MM1. It consists of multiple shocks blended together. The H$_2$ 2.12~$\mu$m and [Fe~{\sc II}] 1.64~$\mu$m line emissions toward the LkH$\alpha$~234 region have been recently investigated by \citet{Oh18} with high-resolution NIR spectral mapping observations. These authors, by studying the line velocity distribution, clearly show that each of the YSOs VLA~3~(A$+$B), VLA~2, and FIRS1-MM1 contributes to the excitation of the H$_2$ emission to southwest, whose very complex structure arises from the overlap of the shocks produced by multiple outflows \citep[][see their Fig.~10]{Oh18}: \ 1)~from VLA~3~(A$+$B), as traced by the [Fe~{\sc II}] jet, detected also in weak H$_2$ emission; \ 2)~from VLA~2, driving low-velocity ($\le$~50~\kms) redshifted collimated H$_2$ emission; \ 3)~from FIRS1-MM1, which drives the H$_2$ jet (the aligned knots 1, 3, 5, and 6 in Fig.~\ref{H2-SW}). 
Conversely, as discussed before, the H$_2$ emission has a much simpler structure to northeast -- a line of well separated knots -- and that agrees with our interpretation that it is excited only by the (parallel) jets from VLA~3A~and~3B. We incidentally note that the isolated knot labeled "A" in Fig.~\ref{H2-all} corresponds well in position with the H$_2$ knot "SP3" studied by \citet[][see their Fig.~10]{Oh18}, and could trace the northeast lobe of the FIRS1-MM1 jet (compare Figs.~\ref{H2-all}~and~\ref{H2-SW}).

In the end, we do not find any clear indication for the northeast lobe of the VLA~2 jet at larger scales in the observed H$_2$ emission pattern (see Fig.~\ref{H2-all}). That is consistent with our discussion of the water maser kinematics $\approx$~300~au northeast of the YSO VLA~2 (see Sect.~\ref{Jsho}), based on which we have postulated the presence of a very dense clump that is (at least) partially blocking the jet propagation. The found agreement between the small and large scales using different jet tracers brings further support to this hypothesis.

\section{Conclusions}
\label{Con}

The POETS survey has shown that VLBI observations of the 22~GHz water masers provide valuable information on the 3D velocity field of protostellar outflows over their launching regions. Intermediate-mass protostars are the closest YSOs that can be reliably investigated with the VLBI water maser technique. As an example, toward the intermediate-mass YSO IRAS~21078$+$5211 we have achieved sufficiently high (sub-au) linear resolution to directly trace the streamlines of the associated MHD DW and accurately determine the launch radii of the wind \citep{Mos22,Mos24}. This result encourages us to pursue our study of intermediate-mass YSOs.

This work focuses on the intermediate-mass star-forming region \targ\ (alias LkH$\alpha$~234), one of the least luminous POETS targets. We combine three different datasets of water maser VLBI observations (two VLBA and one global VLBI), spanning a time of $\approx$~22~yr, with sensitive high-angular-resolution JVLA continuum and LBT H$_2$ 2.12~$\mu$m observations to trace the protostellar outflows at length scales from 10~au to 1~pc, and follow their time evolution, too. Our combined observations allow us to study the protostellar outflows from the intermediate-mass binary system VLA~3A~and~3B, separated by $\approx$~0\pas22, and from VLA~2, an intermediate-mass YSO placed $\approx$~1$^{\prime\prime}$ to northwest of VLA~3. For each of these three YSOs, the elongated structure and spectral index of the radio continuum suggest the presence of a thermal jet. The jets emitted by the YSOs VLA~3A~and~3B are parallel having PAs of \ $\approx$~55\degr\ and \ $\approx$~57\degr, respectively. Toward VLA~2 our VLBA 2011 observations confirm the result from previous VLBA 2001 observations that the water masers are tracing a compact (size $\approx$~400~au) bipolar collimated (PA $\approx$ 70\degr) outflow, i.e., a jet. The two sets of VLBA observations sample the 3D flow velocities sufficiently well to prove that the jet is magneto-centrifugally launched in a MHD DW. We infer launch radii in the range 10--50~au for the streamlines traced by the water masers. Across the three VLBI epochs (2001, 2011, and 2023) the average maser position gets farther from the YSO with time, which can suggest that the jet blows the surrounding dense gas to larger distances from the YSO in the course of time. 

At larger scales of $\sim$~0.1~pc, only the southwest lobe of the VLA~2 jet is traced by the LBT H$_2$ emission. That is consistent with the kinematics of the water masers from the global VLBI 2023 observations, indicating that the jet propagation can be hindered by a very dense clump placed $\approx$~300~au to northeast of VLA~2. Differently from VLA~2, the parallel jets emitted by the nearby YSOs VLA~3A~and~3B can be reliably tracked with the H$_2$ emission at scales of a few 10$^{\prime\prime}$ to both the southwest and the northeast. In particular, northeast of VLA~3 the direction of these two jets crosses a linear chain of spaced H$_2$ knots, which is a clear signature of an episodic jet. Toward VLA~3A there is no clear sign of water maser variability by comparing the VLBA 2011 and global VLBI 2023 observations, both recovering only a few weak water masers. Instead, in VLA~3B, the spatial distribution and intensity of the water masers change significantly between 2011 and 2023, likely reflecting a different state (active in 2011 and quiescent in 2023) of wind ejection. The variable ejection from VLA~3B could be the origin of the episodic jet observed at larger scales.

This work exemplifies the ability of the water maser VLBI observations in determining the 3D velocity field of the YSOs' outflows at length scales of 10--100~au, which is fundamental to investigate their launching mechanism, their interaction with the local environment, and their time behavior. The good correspondence between the findings at these small scales, using the water masers, and those at much larger scales of $\sim$~0.1~pc, from the LBT H$_2$ 2.12~$\mu$m observations, witnesses both the reliability and synergy of these two different outflow tracers.

%
   \bibliographystyle{aa} 
   \bibliography{biblio.bib} 

\begin{appendix}




\onecolumn
\section{Parameters of the H$_2$O masers}

The following two tables list the parameters of the H$_2$O masers from the BeSSeL VLBA and global VLBI observations.

\begin{longtable}{ccccrrrr} 
\caption{\label{wat1} 22.2~GHz H$_2$O maser parameters for \targ\ VLA~2 from the BeSSeL VLBA observations.}\\  
\hline\hline
Feature & Epochs\tablefootmark{a} of & I$_{\rm peak}$ & $V_{\rm LSR}$ & \multicolumn{1}{c}{$\Delta~x$} & \multicolumn{1}{c}{$\Delta~y$} & \multicolumn{1}{c}{$V_{x}$} & \multicolumn{1}{c}{$V_{y}$} \\
Number  & Detection & (Jy beam$^{-1}$) & (km s$^{-1}$) & \multicolumn{1}{c}{(mas)} & \multicolumn{1}{c}{(mas)} & \multicolumn{1}{c}{(km s$^{-1}$)} & \multicolumn{1}{c}{(km s$^{-1}$)} \\
\hline
\endfirsthead
\caption{continued.}\\
\hline\hline
Feature & Epochs\tablefootmark{a} of & I$_{\rm peak}$ & $V_{\rm LSR}$ & \multicolumn{1}{c}{$\Delta~x$} & \multicolumn{1}{c}{$\Delta~y$} & \multicolumn{1}{c}{$V_{x}$} & \multicolumn{1}{c}{$V_{y}$} \\
Number  & Detection & (Jy beam$^{-1}$) & (km s$^{-1}$) & \multicolumn{1}{c}{(mas)} & \multicolumn{1}{c}{(mas)} & \multicolumn{1}{c}{(km s$^{-1}$)} & \multicolumn{1}{c}{(km s$^{-1}$)} \\
\hline
\endhead
\hline
\endfoot
\hline
\endlastfoot
\input{bessel_tab.inp}
\end{longtable} 
\tablefoot{
\\
\tablefoottext{a}{The BeSSeL VLBA epochs are: 1)~May 24, 2011; \ 2)~August 8, 2011; \ 3)~October 30, 2011; \ 4)~November 26, 2011; \ 5)~January 12, 2012; \ 6)~May 14, 2012.} \\
Column~1 gives the feature label number; Column~2 lists the observing epochs at which the feature was detected;
Columns~3~and~4 provide the intensity of the strongest spot
and the intensity-weighted \Vlsr, respectively, averaged over the
observing epochs; Columns~5~and~6 give the position offsets (with
the associated errors) along the RA and DEC axes, relative to feature~\#4, measured at the first epoch of detection; Columns~7~and~8 give the components of the absolute proper motion (with the associated errors) along the RA and DEC axes.\\
The absolute position of the feature~\#4 at the first observing epoch on May 24, 2011, is: 
RA~(J2000) = 21$^{\rm h}$ 43$^{\rm m}$ 6\fs2976, DEC~(J2000) = 66\degree\ 06$^{\prime}$ 55\farcs7579, with an accuracy of \ $\pm$0.5~mas. 
}   

\vspace{1.5cm}

\begin{longtable}{cccrr} 
\caption{\label{wat2} 22.2~GHz H$_2$O maser parameters for \targ\ from the global VLBI observations.}\\  
\hline\hline
Feature &  I$_{\rm peak}$ & $V_{\rm LSR}$ & \multicolumn{1}{c}{$\Delta~x$} & \multicolumn{1}{c}{$\Delta~y$} \\
Number  &  (Jy beam$^{-1}$) & (km s$^{-1}$) & \multicolumn{1}{c}{(mas)} & \multicolumn{1}{c}{(mas)}  \\
\hline
\endfirsthead
\caption{continued.}\\
\hline\hline
Feature &  I$_{\rm peak}$ & $V_{\rm LSR}$ & \multicolumn{1}{c}{$\Delta~x$} & \multicolumn{1}{c}{$\Delta~y$} \\
Number  & (Jy beam$^{-1}$) & (km s$^{-1}$) & \multicolumn{1}{c}{(mas)} & \multicolumn{1}{c}{(mas)}  \\
\hline
\endhead
\hline
\endfoot
\hline
\endlastfoot
\input{GM082_tab.inp}

\end{longtable} 
\tablefoot{
\\
Column~1 gives the feature label number; 
Columns~2~and~3 provide the intensity of the strongest spot
and the intensity-weighted \Vlsr, respectively; Columns~4~and~5 give the position offsets (with
the associated errors) along the RA and DEC axes, relative to feature~\#1.\\
The absolute position of the feature~\#1 at the observing epoch on June 6, 2023, is: 
RA~(J2000) = 21$^{\rm h}$ 43$^{\rm m}$ 06\fs3595, DEC~(J2000) = 66\degree\ 06$^{\prime}$ 55\farcs9551, with an accuracy of \ $\pm$0.5~mas. 
}   

\end{appendix}

\end{document}

%% file: bessel_tab.inp
     1 &              6 &     6.210 &    -6.8 &   372.27$\pm$0.07 &   238.97$\pm$0.05 &    ....\hspace*{0.5cm} &    ....\hspace*{0.5cm}  \\
    2 &              2 &     5.716 &    -8.6 &   114.33$\pm$0.14 &   157.13$\pm$0.14 &    ....\hspace*{0.5cm} &    ....\hspace*{0.5cm}  \\
    3 &              2 &     5.658 &    -8.8 &   113.86$\pm$0.14 &   157.18$\pm$0.10 &    ....\hspace*{0.5cm} &    ....\hspace*{0.5cm}  \\
    4 &    1,2,3,4,5,6 &     2.053 &    -9.6 &     0.00\hspace*{0.4cm} &     0.00\hspace*{0.4cm} &   -7.5$\pm$ 0.4 &  -13.7$\pm$ 0.4  \\
    5 &        1,2,3,5 &     1.950 &    -9.4 &     0.85$\pm$0.10 &    -0.06$\pm$0.10 &   -7.8$\pm$ 1.0 &  -12.6$\pm$ 0.9  \\
    6 &          3,4,5 &     1.622 &   -17.7 &   120.47$\pm$0.16 &   157.61$\pm$0.09 &   15.5$\pm$ 2.3 &   -8.5$\pm$ 2.1  \\
    7 &              2 &     1.585 &    -7.7 &   116.34$\pm$0.06 &   157.28$\pm$0.10 &    ....\hspace*{0.5cm} &    ....\hspace*{0.5cm}  \\
    8 &              2 &     1.332 &    -6.6 &   115.60$\pm$0.12 &   157.07$\pm$0.10 &    ....\hspace*{0.5cm} &    ....\hspace*{0.5cm}  \\
    9 &            1,2 &     1.047 &   -11.5 &   296.71$\pm$0.06 &   213.32$\pm$0.11 &    ....\hspace*{0.5cm} &    ....\hspace*{0.5cm}  \\
   10 &      1,2,3,4,5 &     1.046 &    -4.3 &   373.09$\pm$0.07 &   286.44$\pm$0.10 &   -4.4$\pm$ 0.6 &   10.3$\pm$ 0.7  \\
   11 &    1,2,3,4,5,6 &     0.930 &    -9.5 &    -1.63$\pm$0.10 &     0.62$\pm$0.07 &  -11.5$\pm$ 0.6 &  -11.6$\pm$ 0.3  \\
   12 &              6 &     0.864 &   -12.4 &   130.43$\pm$0.07 &   158.14$\pm$0.05 &    ....\hspace*{0.5cm} &    ....\hspace*{0.5cm}  \\
   13 &      1,2,3,4,5 &     0.853 &    -9.5 &    -0.81$\pm$0.11 &     0.56$\pm$0.07 &  -10.2$\pm$ 0.7 &  -11.7$\pm$ 0.6  \\
   14 &          3,4,5 &     0.766 &   -25.8 &    13.81$\pm$0.14 &   157.21$\pm$0.08 &  -60.0$\pm$ 1.9 &  -11.2$\pm$ 2.4  \\
   15 &              6 &     0.764 &    -6.9 &   373.67$\pm$0.13 &   292.44$\pm$0.20 &    ....\hspace*{0.5cm} &    ....\hspace*{0.5cm}  \\
   16 &      1,2,3,4,5 &     0.725 &   -10.0 &   297.62$\pm$0.10 &   209.17$\pm$0.09 &   31.6$\pm$ 0.9 &   16.6$\pm$ 0.7  \\
   17 &              2 &     0.625 &   -14.2 &   347.79$\pm$0.05 &   206.91$\pm$0.09 &    ....\hspace*{0.5cm} &    ....\hspace*{0.5cm}  \\
   18 &            1,2 &     0.543 &    -9.4 &   335.46$\pm$0.07 &   216.63$\pm$0.07 &    ....\hspace*{0.5cm} &    ....\hspace*{0.5cm}  \\
   19 &              6 &     0.498 &   -14.2 &   -27.36$\pm$0.07 &    57.26$\pm$0.10 &    ....\hspace*{0.5cm} &    ....\hspace*{0.5cm}  \\
   20 &              5 &     0.485 &    -4.6 &   374.12$\pm$0.10 &   289.36$\pm$0.06 &    ....\hspace*{0.5cm} &    ....\hspace*{0.5cm}  \\
   21 &              1 &     0.448 &   -12.4 &   127.47$\pm$0.14 &   149.00$\pm$0.19 &    ....\hspace*{0.5cm} &    ....\hspace*{0.5cm}  \\
   22 &          3,4,5 &     0.384 &   -12.0 &   301.94$\pm$0.17 &   225.44$\pm$0.09 &   35.5$\pm$ 2.6 &   10.8$\pm$ 1.9  \\
   23 &        3,4,5,6 &     0.371 &    -8.7 &   351.68$\pm$0.15 &   205.47$\pm$0.09 &   16.5$\pm$ 1.4 &   -2.6$\pm$ 1.4  \\
   24 &              3 &     0.341 &    -4.4 &   373.31$\pm$0.22 &   288.86$\pm$0.08 &    ....\hspace*{0.5cm} &    ....\hspace*{0.5cm}  \\
   25 &              3 &     0.329 &   -17.1 &   119.92$\pm$0.21 &   157.79$\pm$0.07 &    ....\hspace*{0.5cm} &    ....\hspace*{0.5cm}  \\
   26 &              1 &     0.322 &    -1.9 &   375.28$\pm$0.06 &   282.64$\pm$0.09 &    ....\hspace*{0.5cm} &    ....\hspace*{0.5cm}  \\
   27 &        1,2,3,4 &     0.319 &    -7.8 &   366.66$\pm$0.21 &   236.00$\pm$0.16 &   20.7$\pm$ 1.0 &    5.0$\pm$ 0.9  \\
   28 &        1,3,4,5 &     0.309 &    -9.3 &   346.29$\pm$0.09 &   208.87$\pm$0.09 &   29.4$\pm$ 0.7 &   13.3$\pm$ 0.7  \\
   29 &              5 &     0.287 &   -24.7 &    12.12$\pm$0.05 &   158.71$\pm$0.08 &    ....\hspace*{0.5cm} &    ....\hspace*{0.5cm}  \\
   30 &              6 &     0.268 &   -10.4 &   127.06$\pm$0.12 &   156.71$\pm$0.06 &    ....\hspace*{0.5cm} &    ....\hspace*{0.5cm}  \\
   31 &              6 &     0.258 &    -7.9 &   369.80$\pm$0.10 &   241.50$\pm$0.06 &    ....\hspace*{0.5cm} &    ....\hspace*{0.5cm}  \\
   32 &              6 &     0.258 &   -10.7 &   354.02$\pm$0.23 &   215.34$\pm$0.21 &    ....\hspace*{0.5cm} &    ....\hspace*{0.5cm}  \\
   33 &              6 &     0.257 &    -6.4 &   372.60$\pm$0.10 &   293.06$\pm$0.24 &    ....\hspace*{0.5cm} &    ....\hspace*{0.5cm}  \\
   34 &              2 &     0.230 &    -9.1 &    81.67$\pm$0.12 &   142.22$\pm$0.13 &    ....\hspace*{0.5cm} &    ....\hspace*{0.5cm}  \\
   35 &          4,5,6 &     0.227 &    -8.7 &   352.23$\pm$0.21 &   207.43$\pm$0.20 &   15.9$\pm$ 2.4 &  -10.6$\pm$ 2.8  \\
   36 &              6 &     0.226 &    -8.8 &   307.63$\pm$0.07 &   219.40$\pm$0.08 &    ....\hspace*{0.5cm} &    ....\hspace*{0.5cm}  \\
   37 &          3,4,5 &     0.219 &    -7.1 &   368.06$\pm$0.14 &   238.29$\pm$0.08 &   18.6$\pm$ 2.0 &    1.7$\pm$ 1.8  \\
   38 &        1,3,4,5 &     0.208 &   -10.6 &   344.90$\pm$0.07 &   209.69$\pm$0.10 &   31.8$\pm$ 0.7 &   11.9$\pm$ 0.8  \\
   39 &              5 &     0.188 &    -9.8 &    -0.85$\pm$0.11 &     0.46$\pm$0.09 &    ....\hspace*{0.5cm} &    ....\hspace*{0.5cm}  \\
   40 &              5 &     0.183 &   -12.4 &   -29.70$\pm$0.07 &    59.24$\pm$0.06 &    ....\hspace*{0.5cm} &    ....\hspace*{0.5cm}  \\
   41 &              3 &     0.180 &   -13.0 &   300.65$\pm$0.15 &   229.30$\pm$0.08 &    ....\hspace*{0.5cm} &    ....\hspace*{0.5cm}  \\
   42 &        1,3,5,6 &     0.178 &   -12.6 &   -26.28$\pm$0.08 &    57.73$\pm$0.11 &  -17.2$\pm$ 0.5 &  -10.3$\pm$ 0.6  \\
   43 &          1,2,3 &     0.165 &   -18.5 &   -34.56$\pm$0.08 &    57.69$\pm$0.10 &  -28.3$\pm$ 1.0 &  -12.2$\pm$ 1.2  \\
   44 &              4 &     0.164 &   -26.3 &    12.88$\pm$0.20 &   157.43$\pm$0.22 &    ....\hspace*{0.5cm} &    ....\hspace*{0.5cm}  \\
   45 &              6 &     0.156 &    -8.7 &   373.10$\pm$0.08 &   293.62$\pm$0.12 &    ....\hspace*{0.5cm} &    ....\hspace*{0.5cm}  \\
   46 &          3,4,5 &     0.155 &    -9.9 &     1.24$\pm$0.14 &     0.57$\pm$0.07 &   -6.0$\pm$ 1.9 &  -12.7$\pm$ 1.9  \\
   47 &              1 &     0.153 &   -12.5 &   128.26$\pm$0.10 &   148.68$\pm$0.10 &    ....\hspace*{0.5cm} &    ....\hspace*{0.5cm}  \\
   48 &              6 &     0.153 &   -10.4 &   306.91$\pm$0.28 &   229.53$\pm$0.07 &    ....\hspace*{0.5cm} &    ....\hspace*{0.5cm}  \\
   49 &              2 &     0.153 &   -13.6 &   347.68$\pm$0.06 &   206.14$\pm$0.11 &    ....\hspace*{0.5cm} &    ....\hspace*{0.5cm}  \\
   50 &        1,2,3,5 &     0.149 &    -9.3 &    -2.50$\pm$0.06 &     1.08$\pm$0.08 &  -12.2$\pm$ 0.8 &  -12.5$\pm$ 0.8  \\
   51 &              6 &     0.145 &   -12.8 &   -27.63$\pm$0.11 &    57.98$\pm$0.12 &    ....\hspace*{0.5cm} &    ....\hspace*{0.5cm}  \\
   52 &              1 &     0.143 &   -10.0 &   345.83$\pm$0.11 &   208.74$\pm$0.08 &    ....\hspace*{0.5cm} &    ....\hspace*{0.5cm}  \\
   53 &            4,5 &     0.142 &   -10.8 &   302.53$\pm$0.20 &   228.00$\pm$0.23 &    ....\hspace*{0.5cm} &    ....\hspace*{0.5cm}  \\
   54 &              5 &     0.140 &    -9.8 &     1.05$\pm$0.05 &     0.34$\pm$0.07 &    ....\hspace*{0.5cm} &    ....\hspace*{0.5cm}  \\
   55 &              1 &     0.137 &    -8.2 &   367.57$\pm$0.16 &   236.85$\pm$0.10 &    ....\hspace*{0.5cm} &    ....\hspace*{0.5cm}  \\
   56 &              1 &     0.136 &    -8.2 &   368.32$\pm$0.06 &   237.35$\pm$0.08 &    ....\hspace*{0.5cm} &    ....\hspace*{0.5cm}  \\
   57 &              3 &     0.134 &   -12.4 &   -28.27$\pm$0.15 &    58.54$\pm$0.11 &    ....\hspace*{0.5cm} &    ....\hspace*{0.5cm}  \\
   58 &              5 &     0.128 &   -18.0 &   -36.02$\pm$0.05 &    57.56$\pm$0.08 &    ....\hspace*{0.5cm} &    ....\hspace*{0.5cm}  \\
   59 &              3 &     0.123 &    -3.9 &   372.99$\pm$0.15 &   289.06$\pm$0.10 &    ....\hspace*{0.5cm} &    ....\hspace*{0.5cm}  \\
   60 &              3 &     0.120 &    -9.5 &   301.93$\pm$0.20 &   212.24$\pm$0.08 &    ....\hspace*{0.5cm} &    ....\hspace*{0.5cm}  \\
   61 &              1 &     0.116 &   -15.9 &   -30.57$\pm$0.07 &    52.78$\pm$0.08 &    ....\hspace*{0.5cm} &    ....\hspace*{0.5cm}  \\
   62 &              3 &     0.115 &    -8.1 &   367.70$\pm$0.14 &   296.32$\pm$0.08 &    ....\hspace*{0.5cm} &    ....\hspace*{0.5cm}  \\
   63 &              1 &     0.114 &    -8.9 &   351.47$\pm$0.05 &   190.12$\pm$0.09 &    ....\hspace*{0.5cm} &    ....\hspace*{0.5cm}  \\
   64 &              1 &     0.113 &    -6.3 &   349.85$\pm$0.15 &   188.06$\pm$0.12 &    ....\hspace*{0.5cm} &    ....\hspace*{0.5cm}  \\
   65 &              1 &     0.108 &    -9.8 &   302.08$\pm$0.10 &   210.19$\pm$0.09 &    ....\hspace*{0.5cm} &    ....\hspace*{0.5cm}  \\
   66 &              4 &     0.104 &   -10.7 &   349.64$\pm$0.20 &   211.83$\pm$0.18 &    ....\hspace*{0.5cm} &    ....\hspace*{0.5cm}  \\
   67 &          4,5,6 &     0.102 &    -9.8 &   302.81$\pm$0.19 &   217.88$\pm$0.19 &   15.6$\pm$ 1.0 &    7.0$\pm$ 1.3  \\
   68 &            4,5 &     0.101 &     3.4 &   344.65$\pm$0.20 &   284.47$\pm$0.18 &    ....\hspace*{0.5cm} &    ....\hspace*{0.5cm}  \\
   69 &              1 &     0.092 &   -13.2 &   126.87$\pm$0.10 &   150.22$\pm$0.11 &    ....\hspace*{0.5cm} &    ....\hspace*{0.5cm}  \\
   70 &              1 &     0.085 &   -15.0 &   -29.70$\pm$0.07 &    53.46$\pm$0.10 &    ....\hspace*{0.5cm} &    ....\hspace*{0.5cm}  \\
   71 &              1 &     0.083 &   -16.7 &   -27.20$\pm$0.06 &    55.62$\pm$0.11 &    ....\hspace*{0.5cm} &    ....\hspace*{0.5cm}  \\
   72 &              6 &     0.079 &   -17.8 &   125.38$\pm$0.10 &   157.99$\pm$0.08 &    ....\hspace*{0.5cm} &    ....\hspace*{0.5cm}  \\
   73 &              1 &     0.077 &   -18.2 &   -35.43$\pm$0.08 &    54.55$\pm$0.11 &    ....\hspace*{0.5cm} &    ....\hspace*{0.5cm}  \\
   74 &              5 &     0.076 &   -10.9 &   303.62$\pm$0.07 &   227.80$\pm$0.08 &    ....\hspace*{0.5cm} &    ....\hspace*{0.5cm}  \\
   75 &              1 &     0.073 &   -19.1 &   -34.13$\pm$0.06 &    54.06$\pm$0.11 &    ....\hspace*{0.5cm} &    ....\hspace*{0.5cm}  \\
   76 &              1 &     0.059 &    -8.8 &   370.72$\pm$0.11 &   290.60$\pm$0.16 &    ....\hspace*{0.5cm} &    ....\hspace*{0.5cm}  \\
   77 &              1 &     0.052 &    -8.7 &   367.36$\pm$0.06 &   236.10$\pm$0.14 &    ....\hspace*{0.5cm} &    ....\hspace*{0.5cm}  \\
   78 &              6 &     0.051 &    -9.0 &   308.05$\pm$0.22 &   220.56$\pm$0.18 &    ....\hspace*{0.5cm} &    ....\hspace*{0.5cm}  \\
   79 &              5 &     0.041 &    -9.3 &    -4.55$\pm$0.10 &     1.80$\pm$0.12 &    ....\hspace*{0.5cm} &    ....\hspace*{0.5cm}  \\
   80 &              5 &     0.040 &    -9.0 &   304.57$\pm$0.10 &   213.40$\pm$0.12 &    ....\hspace*{0.5cm} &    ....\hspace*{0.5cm}  \\

%% file: GM082_tab.inp
& & \hspace*{0.7cm}VLA~2 & & \\
\hline
    1 &     2.544 &   -15.4 &     0.00\hspace*{0.4cm} &     0.00\hspace*{0.4cm}  \\
    2 &     0.684 &   -17.2 &  -315.29$\pm$0.08 &  -102.01$\pm$0.06  \\
    3 &     0.391 &   -16.7 &  -315.62$\pm$0.07 &  -102.27$\pm$0.07  \\
    5 &     0.259 &   -15.4 &     0.22$\pm$0.07 &     0.25$\pm$0.04  \\
    6 &     0.167 &   -16.6 &  -314.88$\pm$0.06 &  -101.16$\pm$0.04  \\
    7 &     0.165 &    -6.2 &    -5.09$\pm$0.09 &    14.01$\pm$0.11  \\
   10 &     0.110 &    -8.4 &    14.06$\pm$0.06 &   -73.73$\pm$0.08  \\
   11 &     0.075 &   -14.9 &     0.48$\pm$0.06 &    -0.25$\pm$0.13  \\
   13 &     0.073 &    -9.4 &  -409.36$\pm$0.06 &  -259.71$\pm$0.04  \\
   14 &     0.063 &    -9.1 &   -13.54$\pm$0.07 &    81.42$\pm$0.05  \\
   15 &     0.063 &    -9.5 &   -13.52$\pm$0.06 &    80.97$\pm$0.10  \\
   16 &     0.058 &   -14.4 &     0.55$\pm$0.06 &    -0.51$\pm$0.04  \\
   17 &     0.057 &    -9.6 &   -13.25$\pm$0.07 &    80.21$\pm$0.09  \\
   18 &     0.054 &    -9.3 &  -413.82$\pm$0.07 &  -259.61$\pm$0.05  \\
   19 &     0.048 &    -9.6 &  -408.45$\pm$0.07 &  -259.60$\pm$0.04  \\
   21 &     0.043 &    -9.5 &  -410.08$\pm$0.06 &  -260.04$\pm$0.06  \\
   23 &     0.034 &    -7.5 &    -3.19$\pm$0.06 &    -2.20$\pm$0.04  \\
   25 &     0.033 &    -7.5 &    18.72$\pm$0.06 &   -34.68$\pm$0.05  \\
   29 &     0.023 &   -11.7 &    -0.31$\pm$0.06 &   -15.00$\pm$0.04  \\
   30 &     0.022 &    -7.5 &    -3.30$\pm$0.06 &    -0.95$\pm$0.04  \\
   31 &     0.022 &    -7.1 &    18.55$\pm$0.06 &   -35.56$\pm$0.06  \\
   33 &     0.021 &   -16.8 &  -403.87$\pm$0.06 &  -210.56$\pm$0.04  \\
   34 &     0.021 &    -9.4 &  -408.93$\pm$0.06 &  -259.70$\pm$0.05  \\
   35 &     0.019 &    -8.4 &    -3.37$\pm$0.07 &    -2.23$\pm$0.04  \\
   36 &     0.019 &    -6.3 &    -4.82$\pm$0.07 &    14.25$\pm$0.06  \\
   37 &     0.019 &    -8.8 &  -416.67$\pm$0.07 &  -258.54$\pm$0.06  \\
   39 &     0.018 &   -13.8 &     0.85$\pm$0.06 &    -4.44$\pm$0.06  \\
   40 &     0.018 &    -9.4 &  -411.21$\pm$0.24 &  -259.85$\pm$0.07  \\
   41 &     0.017 &    -4.4 &     1.68$\pm$0.06 &    -5.06$\pm$0.04  \\
   42 &     0.017 &   -14.4 &     0.95$\pm$0.06 &    -0.53$\pm$0.06  \\
   43 &     0.017 &    -8.9 &  -418.74$\pm$0.12 &  -258.31$\pm$0.15  \\
   44 &     0.015 &    -5.0 &    -4.91$\pm$0.07 &    14.58$\pm$0.07  \\
   48 &     0.014 &    -8.9 &  -417.83$\pm$0.08 &  -258.14$\pm$0.20  \\
   49 &     0.012 &    -4.5 &    -4.53$\pm$0.06 &    14.30$\pm$0.05  \\
   50 &     0.012 &    -3.8 &    -3.66$\pm$0.06 &    14.48$\pm$0.05  \\
   51 &     0.012 &    -7.3 &    18.29$\pm$0.07 &   -34.65$\pm$0.05  \\
   52 &     0.011 &   -17.7 &    -0.51$\pm$0.06 &     1.30$\pm$0.05  \\
   54 &     0.011 &    -8.9 &  -114.37$\pm$0.17 &    45.39$\pm$0.09  \\
\hline 
  \multicolumn{5}{c}{VLA~3A}   \\
\hline
   12 &     0.074 &   -31.0 &   667.59$\pm$0.13 &  -985.39$\pm$0.05  \\
   22 &     0.038 &   -31.8 &   666.73$\pm$0.19 &  -985.23$\pm$0.08  \\
   24 &     0.033 &   -29.4 &   670.20$\pm$0.29 &  -985.96$\pm$0.05  \\
   32 &     0.022 &   -29.9 &   669.05$\pm$0.15 &  -985.65$\pm$0.05  \\
\hline 
  \multicolumn{5}{c}{VLA~3B}   \\
\hline
    4 &     0.327 &   -13.1 &   615.54$\pm$0.06 &  -823.59$\pm$0.07  \\
    8 &     0.159 &   -23.0 &   600.24$\pm$0.06 &  -808.57$\pm$0.04  \\
    9 &     0.123 &   -15.4 &   590.91$\pm$0.06 &  -805.29$\pm$0.04  \\
   20 &     0.045 &   -13.8 &   615.31$\pm$0.07 &  -823.94$\pm$0.05  \\
   26 &     0.032 &   -23.3 &   599.90$\pm$0.07 &  -808.65$\pm$0.04  \\
   27 &     0.031 &   -15.6 &   611.24$\pm$0.07 &  -818.41$\pm$0.04  \\
   28 &     0.026 &   -12.8 &   615.76$\pm$0.06 &  -823.36$\pm$0.04  \\
   38 &     0.018 &   -22.7 &   599.65$\pm$0.06 &  -808.68$\pm$0.04  \\
   45 &     0.015 &   -18.0 &   591.21$\pm$0.06 &  -805.28$\pm$0.04  \\
   46 &     0.014 &   -23.1 &   600.33$\pm$0.06 &  -809.04$\pm$0.04  \\
   47 &     0.014 &   -19.7 &   596.74$\pm$0.06 &  -806.12$\pm$0.04  \\
   53 &     0.011 &   -14.6 &   613.26$\pm$0.06 &  -824.58$\pm$0.05  \\

%% file: aa55643-25_astro-ph.bbl
\begin{thebibliography}{39}
\expandafter\ifx\csname natexlab\endcsname\relax\def\natexlab#1{#1}\fi

\bibitem[{{Ageorges} {et~al.}(2010){Ageorges}, {Seifert}, {J{\"u}tte},
  {Knierim}, {Lehmitz}, {Germeroth}, {Buschkamp}, {Polsterer}, {Pasquali},
  {Naranjo}, {Gemperlein}, {Hill}, {Feiz}, {Hofmann}, {Laun}, {Lederer},
  {Lenzen}, {Mall}, {Mand el}, {M{\"u}ller}, {Quirrenbach}, {Sch{\"a}ffner},
  {Storz}, \& {Weiser}}]{Age10}
{Ageorges}, N., {Seifert}, W., {J{\"u}tte}, M., {et~al.} 2010, in Society of
  Photo-Optical Instrumentation Engineers (SPIE) Conference Series, Vol. 7735,
  Ground-based and Airborne Instrumentation for Astronomy III, 77351L

\bibitem[{{Anglada} {et~al.}(2018){Anglada}, {Rodr{\'\i}guez}, \&
  {Carrasco-Gonz{\'a}lez}}]{Ang18}
{Anglada}, G., {Rodr{\'\i}guez}, L.~F., \& {Carrasco-Gonz{\'a}lez}, C. 2018,
  \aapr, 26, 3

\bibitem[{{Bacciotti} {et~al.}(2002){Bacciotti}, {Ray}, {Mundt},
  {Eisl{\"o}ffel}, \& {Solf}}]{Bac02}
{Bacciotti}, F., {Ray}, T.~P., {Mundt}, R., {Eisl{\"o}ffel}, J., \& {Solf}, J.
  2002, \apj, 576, 222

\bibitem[{{Bressan} {et~al.}(2012){Bressan}, {Marigo}, {Girardi}, {Salasnich},
  {Dal Cero}, {Rubele}, \& {Nanni}}]{Bre12}
{Bressan}, A., {Marigo}, P., {Girardi}, L., {et~al.} 2012, \mnras, 427, 127

\bibitem[{{Fuente} {et~al.}(2001){Fuente}, {Neri}, {Mart{\'\i}n-Pintado},
  {Bachiller}, {Rodr{\'\i}guez-Franco}, \& {Palla}}]{Fue01}
{Fuente}, A., {Neri}, R., {Mart{\'\i}n-Pintado}, J., {et~al.} 2001, \aap, 366,
  873

\bibitem[{{Furuya} {et~al.}(2003){Furuya}, {Kitamura}, {Wootten}, {Claussen},
  \& {Kawabe}}]{Fur03}
{Furuya}, R.~S., {Kitamura}, Y., {Wootten}, A., {Claussen}, M.~J., \& {Kawabe},
  R. 2003, \apjs, 144, 71

\bibitem[{{Gaia Collaboration} {et~al.}(2023){Gaia Collaboration}, {Vallenari},
  {Brown}, {Prusti}, {de Bruijne}, {Arenou}, {Babusiaux}, {Biermann},
  {Creevey}, {Ducourant}, {Evans}, {Eyer}, {Guerra}, {Hutton}, {Jordi},
  {Klioner}, {Lammers}, {Lindegren}, {Luri}, {Mignard}, {Panem}, {Pourbaix},
  {Randich}, {Sartoretti}, {Soubiran}, {Tanga}, {Walton}, {Bailer-Jones},
  {Bastian}, {Drimmel}, {Jansen}, {Katz}, {Lattanzi}, {van Leeuwen}, {Bakker},
  {Cacciari}, {Casta{\~n}eda}, {De Angeli}, {Fabricius}, {Fouesneau},
  {Fr{\'e}mat}, {Galluccio}, {Guerrier}, {Heiter}, {Masana}, {Messineo},
  {Mowlavi}, {Nicolas}, {Nienartowicz}, {Pailler}, {Panuzzo}, {Riclet}, {Roux},
  {Seabroke}, {Sordo}, {Th{\'e}venin}, {Gracia-Abril}, {Portell}, {Teyssier},
  {Altmann}, {Andrae}, {Audard}, {Bellas-Velidis}, {Benson}, {Berthier},
  {Blomme}, {Burgess}, {Busonero}, {Busso}, {C{\'a}novas}, {Carry}, {Cellino},
  {Cheek}, {Clementini}, {Damerdji}, {Davidson}, {de Teodoro}, {Nu{\~n}ez
  Campos}, {Delchambre}, {Dell'Oro}, {Esquej}, {Fern{\'a}ndez-Hern{\'a}ndez},
  {Fraile}, {Garabato}, {Garc{\'\i}a-Lario}, {Gosset}, {Haigron}, {Halbwachs},
  {Hambly}, {Harrison}, {Hern{\'a}ndez}, {Hestroffer}, {Hodgkin}, {Holl},
  {Jan{\ss}en}, {Jevardat de Fombelle}, {Jordan}, {Krone-Martins}, {Lanzafame},
  {L{\"o}ffler}, {Marchal}, {Marrese}, {Moitinho}, {Muinonen}, {Osborne},
  {Pancino}, {Pauwels}, {Recio-Blanco}, {Reyl{\'e}}, {Riello}, {Rimoldini},
  {Roegiers}, {Rybizki}, {Sarro}, {Siopis}, {Smith}, {Sozzetti}, {Utrilla},
  {van Leeuwen}, {Abbas}, {{\'A}brah{\'a}m}, {Abreu Aramburu}, {Aerts},
  {Aguado}, {Ajaj}, {Aldea-Montero}, {Altavilla}, {{\'A}lvarez}, {Alves},
  {Anders}, {Anderson}, {Anglada Varela}, {Antoja}, {Baines}, {Baker},
  {Balaguer-N{\'u}{\~n}ez}, {Balbinot}, {Balog}, {Barache}, {Barbato},
  {Barros}, {Barstow}, {Bartolom{\'e}}, {Bassilana}, {Bauchet}, {Becciani},
  {Bellazzini}, {Berihuete}, {Bernet}, {Bertone}, {Bianchi}, {Binnenfeld},
  {Blanco-Cuaresma}, {Blazere}, {Boch}, {Bombrun}, {Bossini}, {Bouquillon},
  {Bragaglia}, {Bramante}, {Breedt}, {Bressan}, {Brouillet}, {Brugaletta},
  {Bucciarelli}, {Burlacu}, {Butkevich}, {Buzzi}, {Caffau}, {Cancelliere},
  {Cantat-Gaudin}, {Carballo}, {Carlucci}, {Carnerero}, {Carrasco},
  {Casamiquela}, {Castellani}, {Castro-Ginard}, {Chaoul}, {Charlot}, {Chemin},
  {Chiaramida}, {Chiavassa}, {Chornay}, {Comoretto}, {Contursi}, {Cooper},
  {Cornez}, {Cowell}, {Crifo}, {Cropper}, {Crosta}, {Crowley}, {Dafonte},
  {Dapergolas}, {David}, {David}, {de Laverny}, {De Luise}, \& {De
  March}}]{Gai23}
{Gaia Collaboration}, {Vallenari}, A., {Brown}, A.~G.~A., {et~al.} 2023, \aap,
  674, A1

\bibitem[{{Goddi} {et~al.}(2011){Goddi}, {Moscadelli}, \& {Sanna}}]{God11a}
{Goddi}, C., {Moscadelli}, L., \& {Sanna}, A. 2011, \aap, 535, L8

\bibitem[{{Greisen}(2003)}]{Gre03}
{Greisen}, E.~W. 2003, in Astrophysics and Space Science Library, Vol. 285,
  Information Handling in Astronomy - Historical Vistas, ed. A.~{Heck}, 109

\bibitem[{{Hawarden} {et~al.}(2001){Hawarden}, {Leggett}, {Letawsky},
  {Ballantyne}, \& {Casali}}]{Haw01}
{Hawarden}, T.~G., {Leggett}, S.~K., {Letawsky}, M.~B., {Ballantyne}, D.~R., \&
  {Casali}, M.~M. 2001, \mnras, 325, 563

\bibitem[{{Hill} {et~al.}(2006){Hill}, {Green}, \& {Slagle}}]{Hil06}
{Hill}, J.~M., {Green}, R.~F., \& {Slagle}, J.~H. 2006, in Society of
  Photo-Optical Instrumentation Engineers (SPIE) Conference Series, Vol. 6267,
  Ground-based and Airborne Telescopes, ed. L.~M. {Stepp}, 62670Y

\bibitem[{{Hollenbach} {et~al.}(2013){Hollenbach}, {Elitzur}, \&
  {McKee}}]{Hol13}
{Hollenbach}, D., {Elitzur}, M., \& {McKee}, C.~F. 2013, \apj, 773, 70

\bibitem[{{Kato} {et~al.}(2011){Kato}, {Fukagawa}, {Perrin}, {Shibai}, {Itoh},
  \& {Ootsubo}}]{Kat11}
{Kato}, E., {Fukagawa}, M., {Perrin}, M.~D., {et~al.} 2011, \pasj, 63, 849

\bibitem[{{Lekht} {et~al.}(2011){Lekht}, {Munitsyn}, {Tolmachev}, \&
  {Krasnov}}]{Lek11}
{Lekht}, E.~E., {Munitsyn}, V.~A., {Tolmachev}, A.~M., \& {Krasnov}, V.~V.
  2011, Astronomy Reports, 55, 857

\bibitem[{{Marvel}(2005)}]{Mar05}
{Marvel}, K.~B. 2005, \aj, 130, 2732

\bibitem[{{Massi} {et~al.}(2003){Massi}, {Lorenzetti}, \& {Giannini}}]{Mas03}
{Massi}, F., {Lorenzetti}, D., \& {Giannini}, T. 2003, \aap, 399, 147

\bibitem[{{Massi} {et~al.}(2000){Massi}, {Lorenzetti}, {Giannini}, \&
  {Vitali}}]{Mas00}
{Massi}, F., {Lorenzetti}, D., {Giannini}, T., \& {Vitali}, F. 2000, \aap, 353,
  598

\bibitem[{{Moscadelli} {et~al.}(2021){Moscadelli}, {Beuther}, {Ahmadi},
  {Gieser}, {Massi}, {Cesaroni}, {S{\'a}nchez-Monge}, {Bacciotti},
  {Beltr{\'a}n}, {Csengeri}, {Galv{\'a}n-Madrid}, {Henning}, {Klaassen},
  {Kuiper}, {Leurini}, {Longmore}, {Maud}, {M{\"o}ller}, {Palau}, {Peters},
  {Pudritz}, {Sanna}, {Semenov}, {Urquhart}, {Winters}, \& {Zinnecker}}]{Mos21}
{Moscadelli}, L., {Beuther}, H., {Ahmadi}, A., {et~al.} 2021, \aap, 647, A114

\bibitem[{{Moscadelli} {et~al.}(2011){Moscadelli}, {Cesaroni}, {Rioja},
  {Dodson}, \& {Reid}}]{Mos11a}
{Moscadelli}, L., {Cesaroni}, R., {Rioja}, M.~J., {Dodson}, R., \& {Reid},
  M.~J. 2011, \aap, 526, A66+

\bibitem[{{Moscadelli} {et~al.}(2007){Moscadelli}, {Goddi}, {Cesaroni},
  {Beltr{\'a}n}, \& {Furuya}}]{Mos07}
{Moscadelli}, L., {Goddi}, C., {Cesaroni}, R., {Beltr{\'a}n}, M.~T., \&
  {Furuya}, R.~S. 2007, \aap, 472, 867

\bibitem[{{Moscadelli} {et~al.}(2024){Moscadelli}, {Oliva}, {Sanna}, {Surcis},
  \& {Bayandina}}]{Mos24}
{Moscadelli}, L., {Oliva}, A., {Sanna}, A., {Surcis}, G., \& {Bayandina}, O.
  2024, \aap, 690, A81

\bibitem[{{Moscadelli} {et~al.}(2016){Moscadelli}, {S{\'a}nchez-Monge},
  {Goddi}, {Li}, {Sanna}, {Cesaroni}, {Pestalozzi}, {Molinari}, \&
  {Reid}}]{Mos16}
{Moscadelli}, L., {S{\'a}nchez-Monge}, {\'A}., {Goddi}, C., {et~al.} 2016,
  \aap, 585, A71

\bibitem[{{Moscadelli} {et~al.}(2022){Moscadelli}, {Sanna}, {Beuther}, {Oliva},
  \& {Kuiper}}]{Mos22}
{Moscadelli}, L., {Sanna}, A., {Beuther}, H., {Oliva}, A., \& {Kuiper}, R.
  2022, Nature Astronomy, 6, 1068

\bibitem[{{Moscadelli} {et~al.}(2019){Moscadelli}, {Sanna}, {Goddi},
  {Krishnan}, {Massi}, \& {Bacciotti}}]{Mos19b}
{Moscadelli}, L., {Sanna}, A., {Goddi}, C., {et~al.} 2019, \aap, 631, A74

\bibitem[{{Moscadelli} {et~al.}(2006){Moscadelli}, {Testi}, {Furuya}, {Goddi},
  {Claussen}, {Kitamura}, \& {Wootten}}]{Mos06}
{Moscadelli}, L., {Testi}, L., {Furuya}, R.~S., {et~al.} 2006, \aap, 446, 985

\bibitem[{{Oh} {et~al.}(2018){Oh}, {Pyo}, {Koo}, {Yuk}, {Kaplan}, {Lee},
  {Sokal}, {Mace}, {Park}, {Lee}, {Park}, {Hwang}, {Kim}, \& {Jaffe}}]{Oh18}
{Oh}, H., {Pyo}, T.-S., {Koo}, B.-C., {et~al.} 2018, \apj, 858, 23

\bibitem[{{Oliva} \& {Kuiper}(2023)}]{OK23b}
{Oliva}, A. \& {Kuiper}, R. 2023, \aap, 669, A81

\bibitem[{{Pickett} {et~al.}(1998){Pickett}, {Poynter}, {Cohen}, {Delitsky},
  {Pearson}, \& {M{\"u}ller}}]{Pick98}
{Pickett}, H.~M., {Poynter}, R.~L., {Cohen}, E.~A., {et~al.} 1998, \jqsrt, 60,
  883

\bibitem[{{Pudritz} {et~al.}(2007){Pudritz}, {Ouyed}, {Fendt}, \&
  {Brandenburg}}]{Pud07}
{Pudritz}, R.~E., {Ouyed}, R., {Fendt}, C., \& {Brandenburg}, A. 2007, in
  Protostars and Planets V, ed. B.~{Reipurth}, D.~{Jewitt}, \& K.~{Keil}, 277

\bibitem[{{Pudritz} \& {Ray}(2019)}]{Pud19}
{Pudritz}, R.~E. \& {Ray}, T.~P. 2019, Frontiers in Astronomy and Space
  Sciences, 6, 54

\bibitem[{{Reid} {et~al.}(2014){Reid}, {Menten}, {Brunthaler}, {Zheng}, {Dame},
  {Xu}, {Wu}, {Zhang}, {Sanna}, {Sato}, {Hachisuka}, {Choi}, {Immer},
  {Moscadelli}, {Rygl}, \& {Bartkiewicz}}]{Rei14}
{Reid}, M.~J., {Menten}, K.~M., {Brunthaler}, A., {et~al.} 2014, \apj, 783, 130

\bibitem[{{Reid} {et~al.}(2009){Reid}, {Menten}, {Brunthaler}, {Zheng},
  {Moscadelli}, \& {Xu}}]{Rei09}
{Reid}, M.~J., {Menten}, K.~M., {Brunthaler}, A., {et~al.} 2009, \apj, 693, 397

\bibitem[{{Sanna} {et~al.}(2010){Sanna}, {Moscadelli}, {Cesaroni}, {Tarchi},
  {Furuya}, \& {Goddi}}]{San10b}
{Sanna}, A., {Moscadelli}, L., {Cesaroni}, R., {et~al.} 2010, \aap, 517, A78+

\bibitem[{{Sanna} {et~al.}(2018){Sanna}, {Moscadelli}, {Goddi}, {Krishnan}, \&
  {Massi}}]{San18}
{Sanna}, A., {Moscadelli}, L., {Goddi}, C., {Krishnan}, V., \& {Massi}, F.
  2018, \aap, 619, A107

\bibitem[{{Sch{\"o}nrich} {et~al.}(2010){Sch{\"o}nrich}, {Binney}, \&
  {Dehnen}}]{Sch10}
{Sch{\"o}nrich}, R., {Binney}, J., \& {Dehnen}, W. 2010, \mnras, 403, 1829

\bibitem[{{Testi} {et~al.}(1999){Testi}, {Palla}, \& {Natta}}]{Tes99}
{Testi}, L., {Palla}, F., \& {Natta}, A. 1999, \aap, 342, 515

\bibitem[{{Torrelles} {et~al.}(2014){Torrelles}, {Curiel}, {Estalella},
  {Anglada}, {G{\'o}mez}, {Cant{\'o}}, {Patel}, {Trinidad}, {Girart},
  {Carrasco-Gonz{\'a}lez}, \& {Rodr{\'\i}guez}}]{Tor14}
{Torrelles}, J.~M., {Curiel}, S., {Estalella}, R., {et~al.} 2014, \mnras, 442,
  148

\bibitem[{{Trinidad} {et~al.}(2004){Trinidad}, {Curiel}, {Torrelles},
  {Rodr{\'{\i}}guez}, {Cant{\'o}}, {G{\'o}mez}, {Patel}, \& {Ho}}]{Tri04}
{Trinidad}, M.~A., {Curiel}, S., {Torrelles}, J.~M., {et~al.} 2004, \apj, 613,
  416

\bibitem[{{Xu} {et~al.}(2013){Xu}, {Li}, {Reid}, {Menten}, {Zheng},
  {Brunthaler}, {Moscadelli}, {Dame}, \& {Zhang}}]{Xu13}
{Xu}, Y., {Li}, J.~J., {Reid}, M.~J., {et~al.} 2013, \apj, 769, 15

\end{thebibliography}
